# Outage Capacity of the Fading Relay Channel in the Low SNR Regime *

A. Salman Avestimehr and David N.C. Tse

Department of Electrical Engineering and Computer Sciences

University of California, Berkeley

Berkeley, CA 94720, USA

Email: { avestime, dtse } @eecs.berkeley.edu

**Abstract**

In slow fading scenarios, cooperation between nodes can increase the amount of diversity for communication. We study the performance limit in such scenarios by analyzing the outage capacity of slow fading relay channels. Our focus is on the low SNR and low outage probability regime, where the adverse impact of fading is greatest but so are the potential gains from cooperation. We showed that while the standard Amplify-Forward protocol performs very poorly in this regime, a modified version we called the Bursty Amplify-Forward protocol is optimal and achieves the outage capacity of the network. Moreover, this performance can be achieved without a priori channel knowledge at the receivers. In contrast, the Decode-Forward protocol is strictly sub-optimal in this regime. Our results directly yield the outage capacity per unit energy of fading relay channels.

*This work is supported by the National Science Foundation under grant CCR-0118784 and by a fellowship from the Vodafone Foundation.

# 1 Introduction

Node cooperation has been shown to be an effective way of providing diversity in wireless fading networks [1, 2, 3]. In the slow fading scenario, once a channel is in deep fade, coding no longer helps to increase the reliability of the transmission. In this situation cooperative transmission can dramatically improve the performance by creating diversity using the antennas available at the other nodes of the network. This observation leads to recent interest in the design and analysis of efficient cooperative transmission protocols.

In this paper, the cooperative diversity scenario is modelled by a slow Rayleigh fading relay channel. There are two regimes of interest that one can look at for this channel : high SNR and low SNR. The design and analysis of cooperative protocols at high SNR have been studied in [3] and [4]. In the high SNR regime, the main performance measure is the *diversity-multiplexing tradeoff* [8], which can be viewed as a high SNR approximation of the outage probability curve. In [3] the authors introduced several simple transmission protocols and analyzed the diversity-multiplexing tradeoff achieved by these schemes. While these schemes extract the maximal available diversity in the channel, they are sub-optimal in terms of achieving the diversity-multiplexing tradeoff. Then in [4] more efficient cooperative transmission protocols were introduced. In particular, they proposed a dynamic decode and forward scheme that achieves the optimal diversity-multiplexing tradeoff in a range of low multiplexing gain.

While at high SNR regime the main challenge is to use the degrees of freedom efficiently, the energy efficiency becomes the important measure in the low SNR regime. Therefore in the low SNR regime we should look for the cooperative schemes that are efficient in the transfer of energy into the network. Moreover based on this intuition the behavior of all protocols can be summarized in how they transfer energy in the network.

In this paper we focus on the outage performance at the low SNR regime. There are two reasons to study the low SNR regime. First, as we will show in section 2, the impact of fading and of diversity on capacity is much more significant in low SNR than high SNR. Second, in energy-limited scenarios, the key performance measure is the maximum number of bits per unit energy that one can communicate for a given $\epsilon$ outage probability. So analogous to [5] one can define the $\epsilon$-outage capacity per unit energy or $\mathbf{C}_\epsilon$. It is easy to show that this capacity is achieved in the low SNR limit and so our results on low SNR outage capacity directly translates to results on the outage capacity per unit cost.

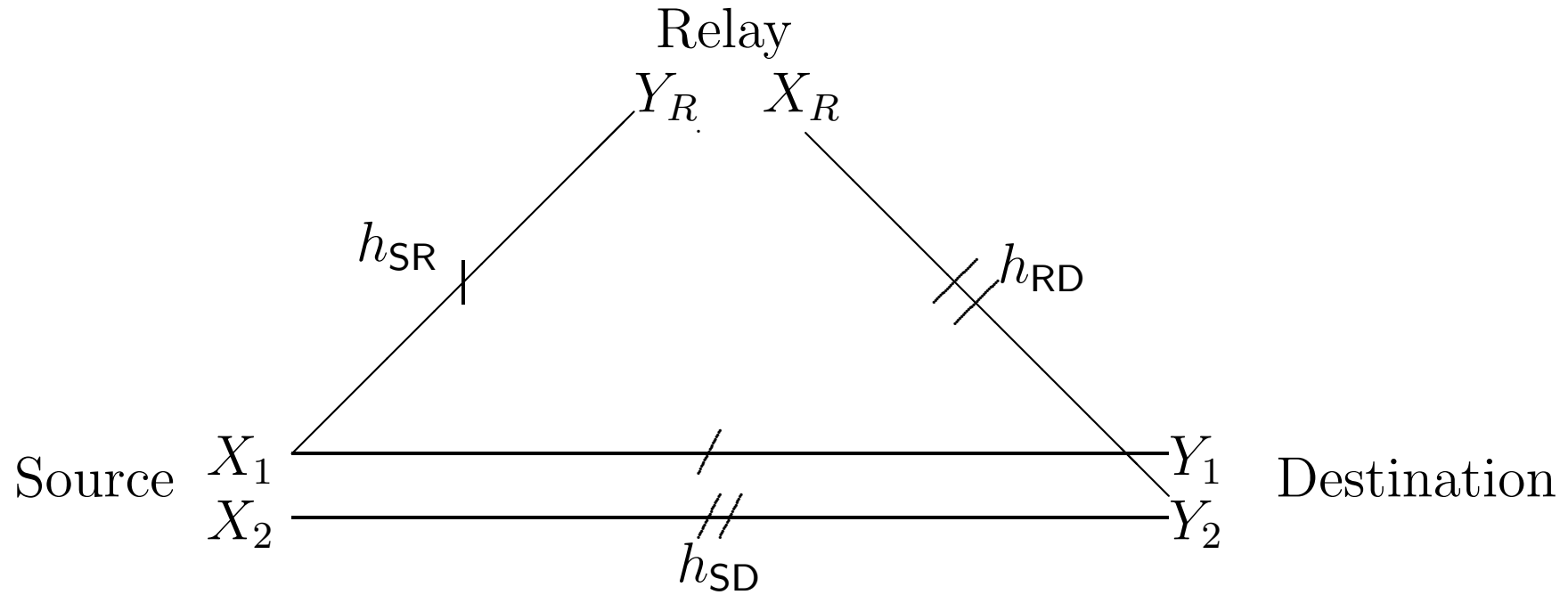

Figure 1: The frequency division communication model that satisfies the half duplex constraint.

We impose a practical constraint on the relay, which is the relay operates on a half-duplex mode and transmits and receives on different frequency bands (so called frequency-division (FD) relay channel). the discrete-time frequency division (FD) model for the fading relay channel with AWGN noise is shown in Figure 1. In order to find the outage capacity we will first use the max-flow min-cut bound to find an upper bound on the $\epsilon$-*outage capacity* of the frequency-division (FD) relay channel. Then we investigate the outage performance of two classes of cooperative protocols: Amplify-Forward (AF) and Decode-Forward (DF) [6]. We show that with AF protocol we get the same outage capacity as when there is no cooperation and only the direct link is used, i.e. no diversity gain. On the other hand with DF protocol we can get full diversity gain. But still there is a gap between the outage capacity of DF protocol and the max-flow min-cut upper bound. Then we investigate the performance of Bursty Amplify-Forward (BAF) protocol, where the source only transmits with a low duty cycle but transmitting at high power when transmitting. We show that, somewhat surprisingly, this simple protocol closes the gap and achieves the optimal outage capacity of the relay channel, in the limit of low SNR and low probability of outage. This leads immediately to the outage capacity per unit cost of FD fading relay channel. The summary of our results is shown in Table 1. In this table $g_{\mathsf{sd}}$, $g_{\mathsf{rd}}$ and $g_{\mathsf{sr}}$ are the variances of channel gains from the source to destination, relay to destination and source to relay respectively and $\mathsf{SNR}_{\mathsf{sd}}$, $\mathsf{SNR}_{\mathsf{rd}}$ and $\mathsf{SNR}_{\mathsf{sr}}$ are the average received SNRs from the source to destination, relay to destination and source to relay. All the noise variances are normalized to be 1. The main results are also stated in the following two theorems.

**Theorem 1.1.** *In the limit of low SNR and low outage probability, the* $\epsilon$-outage *capacity* , $C_{\epsilon_{relay}}$ *of the FD- relay channel(in nats/s) is*

$$C_{\epsilon_{relay}} \approx \sqrt{\frac{2\mathsf{SNR}_{\mathsf{sd}}\mathsf{SNR}_{\mathsf{rd}}\mathsf{SNR}_{\mathsf{sr}}}{\mathsf{SNR}_{\mathsf{rd}} + \mathsf{SNR}_{\mathsf{sr}}}\,\epsilon}. \tag{1}$$

And if we define the $\epsilon$-*outage* capacity per unit energy of the FD- relay channel to be the maximum number of bits that one can transmit with outage probability $\epsilon$, per unit energy spent at the source and unit energy spent at the relay we have

**Theorem 1.2.** *In the limit of low outage probability the* $\epsilon$-outage *capacity per unit energy,* $\mathbf{C}_{\epsilon,\, relay}$, *of the FD- relay channel (in nats/s/J) is*

$$\mathbf{C}_{\epsilon,\, relay} \approx \sqrt{\frac{2g_{\mathsf{sd}}g_{\mathsf{rd}}g_{\mathsf{sr}}}{g_{\mathsf{rd}} + g_{\mathsf{sr}}}\,\epsilon}. \tag{2}$$

The above results assume that the receivers have perfect knowledge of the respective channel gains (but no channel knowledge at the transmitter). As channel estimation is quite challenging in the low SNR regime we ask the following natural question: is channel knowledge crucial in this regime? We show that when neither the transmitters nor the receivers know the channel, the outage capacity is the same as before. An optimal scheme in this case is to use bursty pulse position modulation (PPM) encoding at the source and energy estimation at the destination while the relay just amplifies and forwards the received signal. Therefore we can achieve the same outage performance even in the absence of CSI at the receiver.

In the other extreme we look at the case that the CSI is available at both the transmitter and receiver (full CSI). In this case the source and the relay can beam-form to the destination to obtain better outage performance. To understand how beneficial this additional information can be, we derive the outage capacity. We show that the mixed protocol of bursty amplify-forward + beamforming is the optimal strategy in this case. We also show that for some typical cases the gain from this additional knowledge is small as the source tends to allocate less power for beam-forming and more power to broadcast the information.

| Scenario | Outage Rate(nats/s) |
|---|---|
| Non Cooperative | $\epsilon\ \mathsf{SNR_{sd}}$ |
| Amplify-Forward (AF) | $\epsilon\ \mathsf{SNR_{sd}}$ |
| Decode-Forward (DF) | $\sqrt{\frac{2\mathsf{SNR_{sd}}\mathsf{SNR_{rd}}\mathsf{SNR_{sr}}}{2\mathsf{SNR_{rd}}+\mathsf{SNR_{sr}}}\ \epsilon}$ |
| Bursty Amplify-Forward (BAF) | $\sqrt{\frac{2\mathsf{SNR_{sd}}\mathsf{SNR_{rd}}\mathsf{SNR_{sr}}}{\mathsf{SNR_{rd}}+\mathsf{SNR_{sr}}}\ \epsilon}$ |
| Upper Bound On the Outage Capacity | $\sqrt{\frac{2\mathsf{SNR_{sd}}\mathsf{SNR_{rd}}\mathsf{SNR_{sr}}}{\mathsf{SNR_{rd}}+\mathsf{SNR_{sr}}}\ \epsilon}$ |
| Outage Capacity | $\sqrt{\frac{2\mathsf{SNR_{sd}}\mathsf{SNR_{rd}}\mathsf{SNR_{sr}}}{\mathsf{SNR_{rd}}+\mathsf{SNR_{sr}}}\ \epsilon}$ |
| Outage Capacity per Unit Cost | $\sqrt{\frac{2g_{\mathsf{sd}}g_{\mathsf{rd}}g_{\mathsf{sr}}}{g_{\mathsf{rd}}+g_{\mathsf{sr}}}\ \epsilon}$ |

Table 1: The results on the approximate outage rates (nats/s) at low SNR and low probability of outage $\epsilon$.

# 2 The Effect of Diversity in High vs. Low SNR

In this section we discuss why the effect of diversity is much more significant in low SNR than high SNR. In the case of *slow fading*, where the delay requirement is short compared to the coherence time of the channel, the correct performance measure is the $\epsilon$-*outage capacity* $C_\epsilon$. This is the largest rate of transmission $R(nats/s)$ such that the outage probability $p_{out}(R)$ is less than $\epsilon$. To show the impact of fading, the $\epsilon$-outage capacity of a *point-to-point* Rayleigh fading channel is plotted in Figure 2 as a fraction of the AWGN capacity at the same SNR. It is clear that the impact is much more significant in the low SNR regime. Some simple calculations can explain why. Conditional on a realization of the channel $h$, the fading channel is an AWGN channel with received signal-to-noise ratio equal to $|h|^2\mathsf{SNR}$. Thus

$$\begin{aligned} P_{out_{\mathsf{AWGN}}}(R) &= \mathbb{P}\{\ln(1+|h|^2\mathsf{SNR}) < \mathsf{R} \quad (\text{nats/s})\} \\ &= \mathbb{P}\{|h|^2 < \frac{e^R - 1}{\mathsf{SNR}}\} \end{aligned}$$

If $F$ is the cumulative distribution of $|h|^2$, solving $P_{out_{\mathsf{AWGN}}}(R) = \epsilon$ yields

$$C_\epsilon = \ln(1 + F^{-1}(1-\epsilon)\mathsf{SNR}) \quad (\text{nats/s}) \tag{3}$$

At high SNR we have

$$\begin{aligned} C_\epsilon &\approx \ln \mathsf{SNR} + \ln(\mathsf{F}^{-1}(1-\epsilon)) \quad (\text{nats/s}) \\ &\approx C_{\mathsf{AWGN}} - \ln(\frac{1}{F^{-1}(1-\epsilon)}) \quad (\text{nats/s}), \end{aligned}$$

a constant *difference* irrespective of the SNR. Thus, the relative loss gets smaller at high SNR. At low SNR, on the other hand,

$$\begin{aligned} C_\epsilon &\approx F^{-1}(1-\epsilon)\ \mathsf{SNR} \quad (\text{nats/s}) \\ &\approx F^{-1}(1-\epsilon)\ C_{\mathsf{AWGN}} \end{aligned} \tag{4}$$

But for Rayleigh fading, $|h|^2$, $F(x) = \mathbb{P}\{|h|^2 > x\} = e^{-x}$. Thus for small $\epsilon$

$$\begin{aligned} F^{-1}(1-\epsilon) &= -\ln(1-\epsilon) \\ &\approx \epsilon \end{aligned}$$

This combined with (4) shows that at low SNR, for small outage probability, $\epsilon$, we have

$$C_\epsilon \approx \epsilon\ C_{\mathsf{AWGN}} \tag{5}$$

which is proportional to $\epsilon$ and shows the significant effect of fading at low SNR.

Now we increase the diversity of the channel by having $L$ receive antennas instead of one, each independently Rayleigh faded. The impact of receive diversity on the $\epsilon$-outage capacity for various values of $L$ is plotted in Figure 3. Compared to Figure 2, the dramatic effect of diversity on outage capacity at low SNR can now be seen. For given channel gains $\mathbf{h} := [h_1, \ldots, h_L]^t$, the capacity is $\ln(1 + ||\mathbf{h}||^2 \mathsf{SNR})$. The effective gain $\|\mathbf{h}\|^2$ is $\chi^2$-distributed with $2L$ degrees of freedom. At high SNR the outage probability is given by:

$$p_{out}(R) \approx \frac{(e^R - 1)^L}{L! \mathsf{SNR}^{\mathsf{L}}}. \tag{6}$$

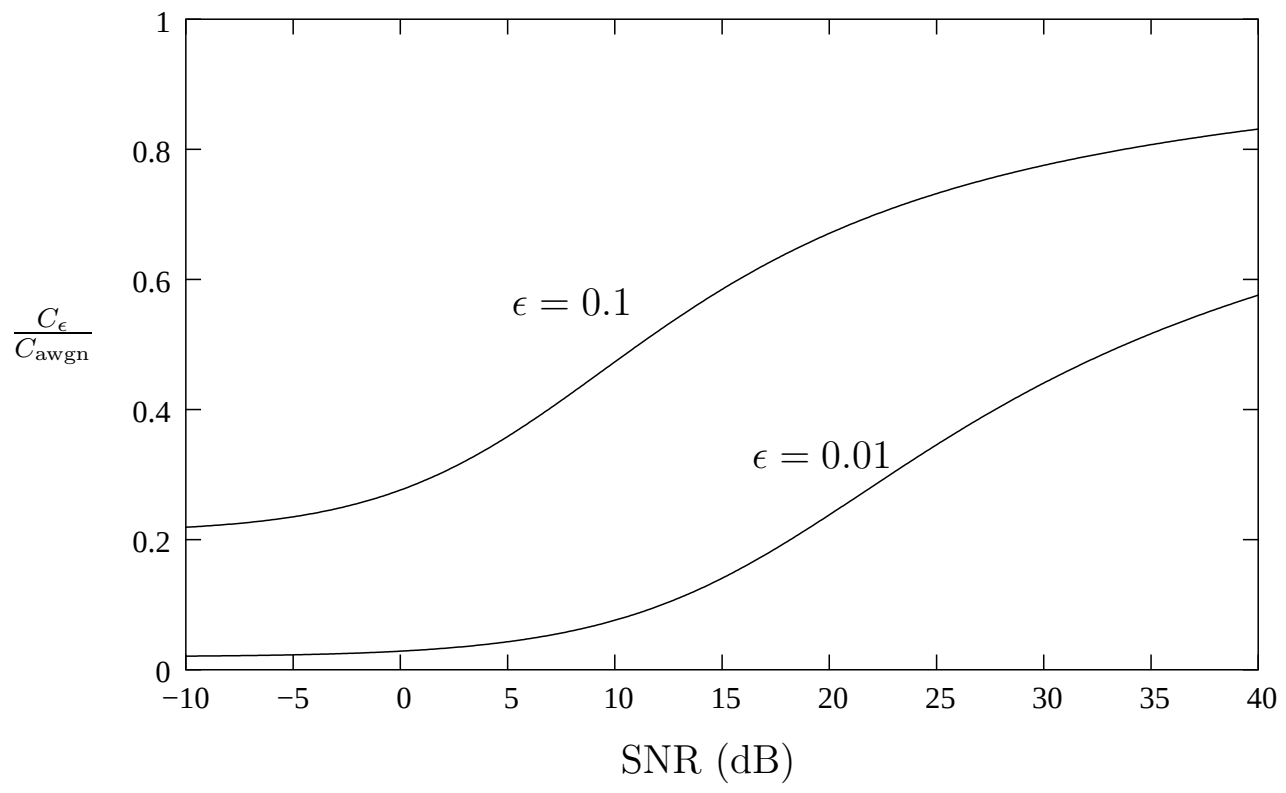

Figure 2: $\epsilon$-outage capacity as a fraction of AWGN capacity under Rayleigh fading, for $\epsilon = 0.1$ and $\epsilon = 0.01$.

Here we see a diversity gain of $L$: the outage probability now decays like $1/\mathsf{SNR}^{\mathsf{L}}$. Let us look at the low-SNR regime. At low SNR and small $\epsilon$, :

$$\begin{aligned} C_\epsilon &\approx F^{-1}(1-\epsilon)\mathsf{SNR} && (7) \\ &\approx (L!)^{\frac{1}{L}}(\epsilon)^{\frac{1}{L}}\mathsf{SNR} \quad \text{nats/s} && (8) \end{aligned}$$

and the loss with respect to the AWGN capacity is by a factor of $\epsilon^{1/L}$ rather than by $\epsilon$ when there is no diversity. For example at low SNR and at $\epsilon = 0.01$, for $L = 1$, the outage capacity i is only 1% of the AWGN capacity and for $L = 2$ it is dramatically increased to 14% of the AWGN capacity. Note that in this regime, the diversity $L$ is reflected in the exponent of $\epsilon$ in the outage capacity.

# 3 Model

In this paper we consider a simple relay network consisting of a source (S), a relay (R) and a destination (D). We impose a practical constraint on the relay that does not allow the relay to receive and transmit signals simultaneously at the same time and the same frequency band, known as the half-duplex constraint. There are two major models in the literature that satisfy this constraint: fixed and random division strategies. In the fixed division strategy the relay receives and transmits data on different frequency-bands/time-slots (frequency division/time division). In the random division strategy the relay randomly decides to listen to data or transmit at each time slot. In this paper we consider the fixed division strategy and the discrete-time frequency division (FD) model for the fading relay channel with AWGN noise is shown in Figure 1. We focus on the case that the channel from the source to the relay and from the relay to the destination is split into two bands. The path gains $h_{\mathsf{sd}}$, $h_{\mathsf{rd}}$ and $h_{\mathsf{sr}}$ are subject to independent Rayleigh fading with variances $g_{\mathsf{sd}}$, $g_{\mathsf{rd}}$ and $g_{\mathsf{sr}}$ respectively. The received signal at the relay at time $i \geq 1$ is

$$Y_{R_i} = h_{\mathsf{sr}} X_{1_i} + Z_{R_i}$$

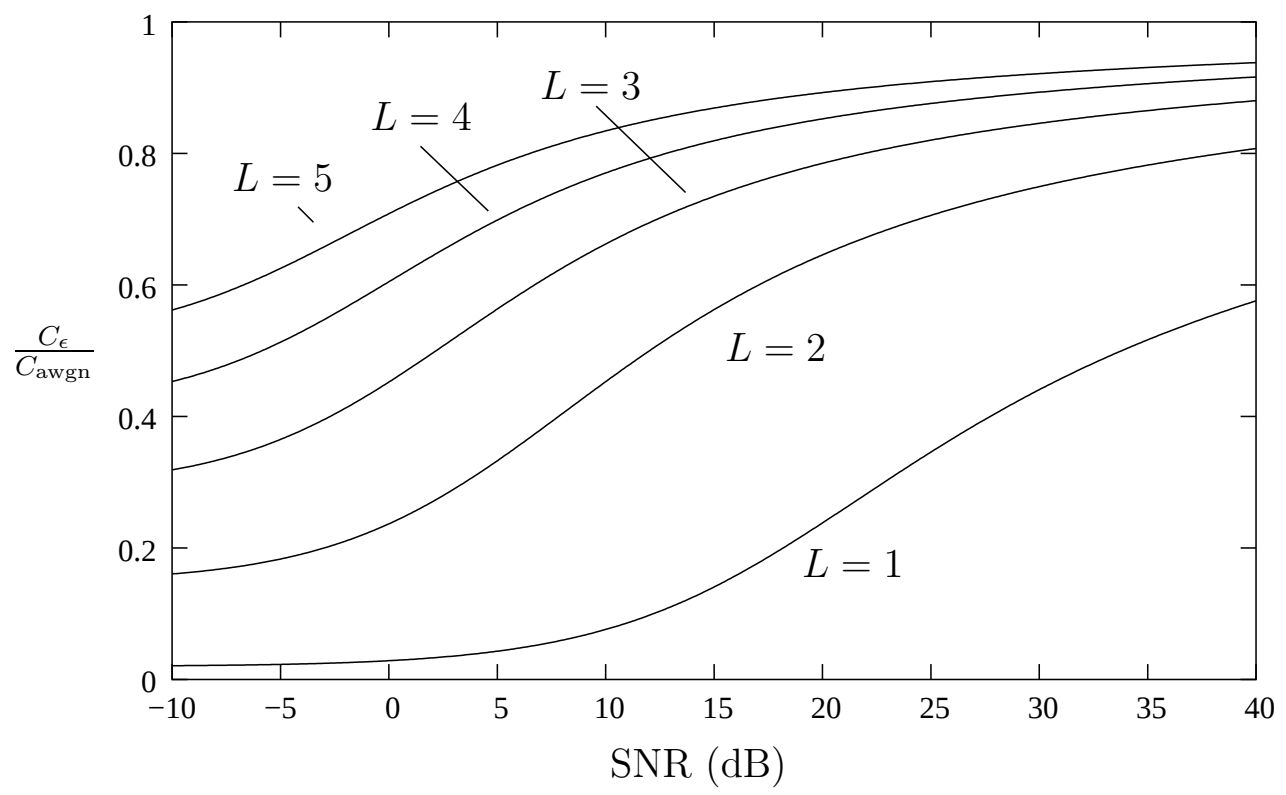

Figure 3: $\epsilon$-outage capacity with $L$-fold receive diversity, as a fraction of the AWGN capacity $\ln(1 + L\,\mathsf{SNR})$, for $\epsilon = 0.01$ and different $L$.

The received signals at time $i$ at the destination from the first and the second frequency bands are denoted by $Y_{1_i}$ and $Y_{2_i}$ respectively, where $Y_{1_i} = h_{\mathsf{sd}} X_{1_i} + Z_{1_i}$ and $Y_{2_i} = h_{\mathsf{rd}} X_{R_i} + h_{\mathsf{sd}} X_{2_i} + Z_{2_i}$. Also $\{Z_{R_i}\}$, $\{Z_{1_i}\}$ and $\{Z_{2_i}\}$ are assumed to be independent (over time and with each other) $\mathcal{CN}(0, 1)$ noises. An average transmitted power constraint equal to $P$ at both the source and the relay is assumed. We also define $\mathsf{SNR} := \mathsf{P}/1$ as the SNR per (complex) degree-of-freedom. Therefore the average received SNRs from the source to destination ($\mathsf{SNR}_{\mathsf{sd}}$), relay to destination ($\mathsf{SNR}_{\mathsf{rd}}$) and source to relay ($\mathsf{SNR}_{\mathsf{sr}}$) are equal to

$$\mathsf{SNR}_{\mathsf{sd}} = g_{\mathsf{sd}}\mathsf{SNR}$$

$$\mathsf{SNR}_{\mathsf{rd}} = g_{\mathsf{sr}}\mathsf{SNR}$$

$$\mathsf{SNR}_{\mathsf{sr}} = g_{\mathsf{rd}}\mathsf{SNR}$$

We consider the *slow fading* situation where the delay requirement is short compared to the coherence time of the channel. Thus we can assume that the channel gains are random but fixed for all time. We also assume that the relay knows channel gain $h_{\mathsf{sr}}$ and the destination knows the channel gains $h_{\mathsf{sd}}$ and $h_{\mathsf{rd}}$. In this paper we compute the mutual information and rates in nats/s.

# 4 The Outage Capacity of the Relay Channel

## 4.1 The Upper Bound on the Outage Capacity of the FD Relay Channel

In this section we find an upper bound on the $\epsilon - \mathit{outage}$ capacity of the FD relay channel in the limit of low SNR and low probability of outage. The bound is based on the general max-flow min-cut bound for the networks [9]. In the network shown in Figure 1 we have two cuts between the source and the destination: the broadcast cut and the multiple access cut. Also there are two frequency bands in the model: the relay listens on one frequency band and transmits on the other, let the fraction of the bandwidth that relay allocates to listen to source be $\alpha$. Also source should decide on the amount of energy it is going to pour into each frequency band, Let the source use power $P_1$ in the first frequency band and $P_2$ in the other. In order to satisfy the power constraint $P$ at the source we should have:

$$\alpha P_1 + (1-\alpha)P_2 \leq P \tag{9}$$

By using the max-flow min-cut upper bound for a fixed realization of the channels, $h_{\mathsf{sd}}$, $h_{\mathsf{rd}}$ and $h_{\mathsf{sr}}$, and the fact that the maximum average mutual information in the bound can be obtained by independent complex Gaussian random variables that are also independent of the channel gains and maximizing over the choices of $\alpha$, $P_1$ and $P_2$ we have,

$$\begin{aligned}
C_{relay}(h_{\mathsf{sr}}, h_{\mathsf{sd}}, h_{\mathsf{rd}}) &\leq \max_{\substack{0\leq\alpha\leq 1, P_1, P_2\\ \alpha P_1+(1-\alpha)P_2=P}} \min\{\alpha \ln(1+(|h_{\mathsf{sd}}|^2+|h_{\mathsf{sr}}|^2)\ P_1)+(1-\alpha)\ln(1+|h_{\mathsf{sd}}|^2 P_2) \\
&\qquad , \alpha\ln(1+|h_{\mathsf{sd}}|^2 P_1)+(1-\alpha)\ln(1+|h_{\mathsf{rd}}|^2 P+|h_{\mathsf{sd}}|^2 P_2)\} \\
&\leq \max_{\substack{0\leq\alpha\leq 1, P_1, P_2\\ \alpha P_1+(1-\alpha)P_2=P}} \min\{\alpha|h_{\mathsf{sr}}|^2 P_1+|h_{\mathsf{sd}}|^2(\alpha P_1+(1-\alpha)P_2) \\
&\qquad , |h_{\mathsf{sd}}|^2(\alpha P_1+(1-\alpha)P_2)+|h_{\mathsf{rd}}|^2(1-\alpha)P\} \\
&= P \max_{\substack{0\leq\beta\leq 1\\ \alpha P_1=\beta P}} \{|h_{\mathsf{sd}}|^2+\min(|h_{\mathsf{rd}}|^2, \beta|h_{\mathsf{sr}}|^2)\} \qquad (10) \\
&= (|h_{\mathsf{sd}}|^2+\min\{|h_{\mathsf{rd}}|^2, |h_{\mathsf{sr}}|^2\})\mathsf{SNR} \qquad (11)
\end{aligned}$$

Although (11) is an upper bound on the max-flow min-cut bound for the FD-relay channel, but as we move to low SNR regime this bound gets tight to the max-flow min-cut bound (because $\ln(1+x) \approx x$ for small $x$). It is very important to note that (10) is maximized when $\beta = 1$, which means that source is not allocating any energy to the second frequency band. This is due to the fact that at low SNR regime we are not degrees of freedom limited therefore the optimal strategy is to allocate all energy to the frequency band that both the relay and the destination have access to. The other important fact is that the important parameter is the amount of energy that is transmitted in each frequency band not how to divide the frequency band.

Also the upper bound shown in (11) can be viewed as the max-flow min-cut bound on the energy flow from the source to the destination therefore one can conclude that any optimal scheme that provides a rate close to the max-flow min-cut bound should be an energy efficient scheme.

Now we can use this bound to get the following upper bound on the outage capacity of the FD-relay channel:

**Theorem 4.1.** *The* $\epsilon$-outage *capacity* , $C_{\epsilon_{relay}}$, *of the FD- relay channel (in nats/s) satisfies*

$$\lim_{\substack{\epsilon\to 0\\ \mathsf{SNR}\to 0}} \frac{C_{\epsilon_{relay}}}{\sqrt{\epsilon}\,\mathsf{SNR}} \leq \sqrt{\frac{2g_{\mathsf{sd}}g_{\mathsf{rd}}g_{\mathsf{sr}}}{g_{\mathsf{rd}}+g_{\mathsf{sr}}}} \tag{12}$$

*Proof.* The max-flow min-cut bound shown in (11) implies that

$$\begin{aligned} P_{out_{relay}}(R) &= \mathbb{P}\{C_{relay}(h_{\mathsf{sr}}, h_{\mathsf{sd}}, h_{\mathsf{rd}}) < R\} \\ &\geq \mathbb{P}\{\mathsf{SNR}(|\mathsf{h}_{\mathsf{sd}}|^2 + \min(|\mathsf{h}_{\mathsf{rd}}|^2, |\mathsf{h}_{\mathsf{sr}}|^2)) < \mathsf{R}\} \\ &= \mathbb{P}\{|h_{\mathsf{sd}}|^2 + \min(|h_{\mathsf{rd}}|^2, |h_{\mathsf{sr}}|^2) < \frac{R}{\mathsf{SNR}}\} \end{aligned}$$

Now $\min(|h_{\mathsf{rd}}|^2, |h_{\mathsf{sr}}|^2)$ is another exponential random variable with mean $\frac{g_{\mathsf{rd}}g_{\mathsf{sr}}}{g_{\mathsf{rd}}+g_{\mathsf{sr}}}$ also $\epsilon \to 0$ implies that $\frac{R}{\mathsf{SNR}} \to 0$, thus from Lemma A.1 we have

$$\begin{aligned} \lim_{\substack{\mathsf{SNR}\to 0\\ \frac{R}{\mathsf{SNR}}\to 0}} \frac{p_{out_{relay}}(R)}{\left(\frac{R}{\mathsf{SNR}}\right)^2} &\geq \lim_{\substack{\mathsf{SNR}\to 0\\ \frac{R}{\mathsf{SNR}}\to 0}} \frac{\mathbb{P}\{|h_{\mathsf{sd}}|^2 + \min(|h_{\mathsf{rd}}|^2, |h_{\mathsf{sr}}|^2) < \frac{R}{\mathsf{SNR}}\}}{\left(\frac{R}{\mathsf{SNR}}\right)^2} \\ &= \frac{1}{2 \times g_{\mathsf{sd}} \times \frac{g_{\mathsf{rd}}g_{\mathsf{sr}}}{g_{\mathsf{rd}}+g_{\mathsf{sr}}}} = \frac{g_{\mathsf{rd}}+g_{\mathsf{sr}}}{2g_{\mathsf{sd}}g_{\mathsf{rd}}g_{\mathsf{sr}}} \end{aligned}$$

Thus

$$\lim_{\substack{\epsilon\to 0\\ \mathsf{SNR}\to 0}} \frac{C_{\epsilon_{relay}}}{\sqrt{\epsilon}\,\mathsf{SNR}} \leq \sqrt{\frac{2g_{\mathsf{sd}}g_{\mathsf{rd}}g_{\mathsf{sr}}}{g_{\mathsf{rd}}+g_{\mathsf{sr}}}} \tag{13}$$

□

It is instructive to compare this upper bound to the outage capacity of a $2 \times 1$ MISO channel which is also another upper bound on the outage capacity of FD-fading relay channel. For the MISO channel we have

$$\begin{aligned} \mathbb{P}_{\mathsf{out,MISO}} &= \mathbb{P}\{\ln(1 + (|h_{\mathsf{sd}}|^2 + |h_{\mathsf{rd}}|^2)\mathsf{SNR}) < \mathsf{R}\} \\ &= \mathbb{P}\{|h_{\mathsf{sd}}|^2 + |h_{\mathsf{rd}}|^2 < \frac{e^R - 1}{\mathsf{SNR}}\} \end{aligned}$$

therefore by lemma, ( A.1) ,

$$\lim_{\substack{\mathsf{SNR}\to 0\\ \frac{R}{\mathsf{SNR}}\to 0}} \frac{p_{\mathsf{out_{MISO}}}(R)}{\left(\frac{R}{\mathsf{SNR}}\right)^2} = \frac{1}{2g_{\mathsf{sd}}g_{\mathsf{rd}}} \tag{14}$$

which shows that

$$\lim_{\substack{\epsilon \to 0 \\ \mathsf{SNR} \to 0}} \frac{C_{\epsilon_{\mathsf{MISO}}}}{\sqrt{\epsilon}\,\mathsf{SNR}} \leq \sqrt{2 g_{\mathsf{sd}} g_{\mathsf{rd}}} \tag{15}$$

Now by comparing (12) to (15) we notice that the max-flow min-cut bound is generally tighter than the more straight forward MISO upper bound (because $\frac{g_{\mathsf{sr}}}{g_{\mathsf{sr}}+g_{\mathsf{rd}}} \leq 1$).

The derived upper bound also shows that up to the first order approximation the outage capacity of the relay channel, $C_{\epsilon_{relay}}$ is upperbounded by

$$\sqrt{\frac{2 g_{\mathsf{sd}} g_{\mathsf{rd}} g_{\mathsf{sr}}}{g_{\mathsf{rd}} + g_{\mathsf{sr}}}\,\epsilon}\,\mathsf{SNR} = \sqrt{\frac{2\mathsf{SNR}_{\mathsf{sd}}\mathsf{SNR}_{\mathsf{rd}}\mathsf{SNR}_{\mathsf{sr}}}{\mathsf{SNR}_{\mathsf{rd}} + \mathsf{SNR}_{\mathsf{sr}}}\,\epsilon} \tag{16}$$

Now having this upper bound on the outage capacity of FD-fading relay channel in the low SNR and low outage probability regime, one might wonder how close the outage capacity can be to this bound. To answer this question we first analyze two commonly used schemes for the relay channel: decode-forward and amplify-forward schemes. We show that although the achievable outage rate of these two schemes is not the same as max-flow min-cut upper bound, they suggest a natural scheme called bursty amplify-forward (BAF) scheme that should be used the achieve the upper bound and thereby we establish the outage capacity.

## 4.2 The Achievable Outage Rate of Decode-Forward Protocol

In this section we look at the decode-forward strategy. In this scheme first, the message is broadcasted to both the destination and the relay from the source. Then the relay tries to decode the message. If the relay is successful in decoding the message, then re-transmits it by using repetition coding (while the source is silent), otherwise remains silent. It is easy to verify that the maximum mutual information achieved with this strategy is

$$I_{DF} = \begin{cases} \ln(1 + |h_{\mathsf{sd}}|^2\mathsf{SNR}) & |h_{\mathsf{sr}}|^2 < (e^R - 1)/\mathsf{SNR} \\ \ln(1 + (|h_{\mathsf{sd}}|^2 + |h_{\mathsf{rd}}|^2)\mathsf{SNR}) & |h_{\mathsf{sr}}|^2 > (e^R - 1)/\mathsf{SNR} \end{cases} \tag{17}$$

where the first case is related to the time that relay is not able to decode so the message is transmitted only through the direct link between the source and the destination, therefore we can easily compute the rate that this scheme can provide for outage probability of $\epsilon$:

**Theorem 4.2.** *For the achievable $\epsilon - outage$ rate of decode-forward scheme we have*

$$\lim_{\substack{\epsilon\to 0\\ \mathsf{SNR}\to 0}} \frac{R_{\epsilon_{DF}}}{\sqrt{\epsilon}\,\mathsf{SNR}} = \sqrt{\frac{2g_{\mathsf{sd}}g_{\mathsf{rd}}g_{\mathsf{sr}}}{2g_{\mathsf{rd}}+g_{\mathsf{sr}}}} \tag{18}$$

*Proof.*

$$\begin{aligned} P_{out_{DF}}(R) &= \mathbb{P}\{I_{DF} < R\} \\ &= \mathbb{P}\{|g|^2 < \frac{e^R-1}{\mathsf{SNR}}\}\mathbb{P}\{|h_{\mathsf{sd}}|^2 < \frac{e^R-1}{\mathsf{SNR}}\} \\ &\quad + \mathbb{P}\{|h_{\mathsf{sr}}|^2 > \frac{e^R-1}{\mathsf{SNR}}\}\mathbb{P}\{|h_{\mathsf{sd}}|^2 + |h_{\mathsf{rd}}|^2 < \frac{e^R-1}{\mathsf{SNR}}\} \end{aligned}$$

and as $\frac{e^R-1}{\mathsf{SNR}} \geq \frac{R}{\mathsf{SNR}}$, $\epsilon \to 0$ implies that $\frac{R}{\mathsf{SNR}} \to 0$ thus

$$\begin{aligned} \lim_{\substack{\epsilon\to 0\\ SNR\to 0}} \frac{P_{out_{DF}}}{\left(\frac{R}{\mathsf{SNR}}\right)^2} &= \lim_{\substack{\mathsf{SNR}\to 0\\ \frac{R}{\mathsf{SNR}}\to 0}} \left(\frac{\mathbb{P}\{|h_{\mathsf{sr}}|^2 < \frac{e^R-1}{\mathsf{SNR}}\}}{\frac{R}{\mathsf{SNR}}} \times \frac{\mathbb{P}\{|h_{\mathsf{sd}}|^2 < \frac{e^R-1}{\mathsf{SNR}}\}}{\frac{R}{\mathsf{SNR}}}\right) \\ &\quad + \lim_{\substack{\mathsf{SNR}\to 0\\ \frac{R}{\mathsf{SNR}}\to 0}} \left(\mathbb{P}\{|h_{\mathsf{sr}}|^2 > \frac{e^R-1}{\mathsf{SNR}}\} \times \frac{\mathbb{P}\{|h_{\mathsf{sd}}|^2 + |h_{\mathsf{rd}}|^2 < \frac{e^R-1}{\mathsf{SNR}}\}}{\left(\frac{R}{\mathsf{SNR}}\right)^2}\right) \\ &= g_{\mathsf{sr}}^{-1} \times g_{\mathsf{sd}}^{-1} + 1 \times \frac{1}{2g_{\mathsf{sd}}g_{\mathsf{rd}}} = \frac{2g_{\mathsf{rd}}+g_{\mathsf{sr}}}{2g_{\mathsf{sd}}g_{\mathsf{rd}}g_{\mathsf{sr}}} \end{aligned}$$

the last equality follows from Lemma A.1 and some calculations. Thus for the achievable $\epsilon - outage$ rate of DF we have

$$\lim_{\substack{\epsilon\to 0\\ \mathsf{SNR}\to 0}} \frac{R_{\epsilon_{DF}}}{\sqrt{\epsilon}\,\mathsf{SNR}} = \sqrt{\frac{2g_{\mathsf{sd}}g_{\mathsf{rd}}g_{\mathsf{sr}}}{2g_{\mathsf{rd}}+g_{\mathsf{sr}}}}$$

□

This theorem shows that at low SNR and for small outage probability, $\epsilon$, the achievable outage rate of decode-forward protocol is

$$R_{\epsilon_{DF}} \approx \sqrt{\frac{2g_{\mathsf{sd}}g_{\mathsf{rd}}g_{\mathsf{sr}}}{2g_{\mathsf{rd}}+g_{\mathsf{sr}}}}\,\epsilon\,\mathsf{SNR} = \sqrt{\frac{2\mathsf{SNR}_{\mathsf{sd}}\mathsf{SNR}_{\mathsf{rd}}\mathsf{SNR}_{\mathsf{sr}}}{2\mathsf{SNR}_{\mathsf{rd}}+\mathsf{SNR}_{\mathsf{sr}}}\,\epsilon} \tag{19}$$

This compared to the upper bound on the outage capacity (12) shows that DF protocol can not achieve the max-flow min-cut upper bound. In fact the ratio between the achievable rate of DF protocol and the upper bound on the outage capacity is $\sqrt{\frac{g_{\mathsf{rd}}+g_{\mathsf{sr}}}{2g_{\mathsf{rd}}+g_{\mathsf{sr}}}} = \sqrt{\frac{g_{\mathsf{rd}}/g_{\mathsf{sr}}+1}{2g_{\mathsf{rd}}/g_{\mathsf{sr}}+1}}$. The first thing to note is that this ratio is a number between $\sqrt{\frac{1}{2}}$ and 1. This ratio is closer to $\sqrt{\frac{1}{2}}$ when the relay is closer to the destination than the source ($g_{\mathsf{rd}}/g_{\mathsf{sr}} > 1$). This follows the reason that more often the attempt to fully decode the message will not be successful in this case. On the other hand when the source is closer to the source than the destination ($g_{\mathsf{rd}}/g_{\mathsf{sr}} < 1$) the relay is able to decode the whole message most of the time and therefore this ratio is closer to 1.

It is interesting to note that in the case of low SNR, even if the relay uses more sophisticated coding rather than repetition coding (for example Slepian-Wolf coding [6]) we can not improve the outage rate of DF protocol. The reason is that at low SNR repetition coding is optimal. Also, in [4] the achievable outage rate of the dynamic-decode-forward scheme was analyzed at high SNR. In this scheme the source transmits the data over the whole time slot and the relay listens until it is able to decode. Once it is able to decode, it helps the transmission. This protocol was shown to be optimal at high SNR, in the sense of diversity-multiplexing trade-off, for the outage performance of the relay channel for some range of multiplexing gain. But at low SNR there is no hope of getting better rate by using the dynamic-decode-forward scheme. The reason is that at low SNR we are not degree-of-freedom limited so the performance of the dynamic-decode-forward scheme is the same as the decode-forward scheme.

One way to interpret why decode-forward protocol does not achieve the max-flow min-cut bound is to analyze how it transfers the energy in the network. Relay may or may not be able to decode and forward the data to the destination and this depends on the channel from the source to the itself. Therefore the energy transferred by the relay is discontinues on the channel gain from the source to the relay. However the max-flow min-cut bound (11) shows that the optimal energy transfer should be continuous in all channel gains. This suggests trying other schemes where the relay behavior is some sort of continuous, one example would be amplify-forward which we analyze in the next section.

## 4.3 The Achievable Outage Rate of Amplify-Forward Protocol

In this strategy first the source broadcasts the message to both the relay and the destination. Then the relay re-scales its received signal to satisfy the power constraint and transmits it to the destination. It is easy to show that for a given realization of the channels $h_{\mathsf{sd}}$, $h_{\mathsf{rd}}$ and $h_{\mathsf{sr}}$ the maximum average mutual information in nats between the input and the two outputs (received signals from the source and the relay at the destination), achieved by i.i.d complex Gaussian inputs is

$$I_{AF} = \ln\left(1 + \mathsf{SNR}(|\mathsf{h}_{\mathsf{sd}}|^2 + \frac{|\mathsf{h}_{\mathsf{rd}}|^2|\mathsf{h}_{\mathsf{sr}}|^2\mathsf{SNR}}{|\mathsf{h}_{\mathsf{rd}}|^2\mathsf{SNR} + |\mathsf{h}_{\mathsf{sr}}|^2\mathsf{SNR} + 1})\right) \tag{20}$$

Therefore we can easily compute the rate that this scheme can provide for outage probability of $\epsilon$:

**Theorem 4.3.** *For the achievable $\epsilon -$ outage rate of amplify-forward scheme we have*

$$\lim_{\substack{\epsilon\to 0\\ \mathsf{SNR}\to 0}} \frac{R_{\epsilon_{AF}}}{\sqrt{\epsilon}\,\mathsf{SNR}} = 0 \tag{21}$$

*and if* $\epsilon > \mathsf{SNR}$

$$\lim_{\substack{\epsilon\to 0\\ \mathsf{SNR}\to 0\\ \epsilon > SNR}} \frac{R_{\epsilon_{AF}}}{\epsilon\,\mathsf{SNR}} = g_{\mathsf{sd}} \tag{22}$$

*Proof.* As the proof of the first and the second part are very similar here we only prove the first part of the Theorem.

$$\begin{aligned}
P_{out_{AF}}(R) &= \mathbb{P}\{I_{AF} < R\} \\
&\geq \mathbb{P}\{|h_{\mathsf{sd}}|^2 + \frac{|h_{\mathsf{rd}}|^2|h_{\mathsf{sr}}|^2\mathsf{SNR}}{|h_{\mathsf{rd}}|^2\mathsf{SNR} + |\mathsf{h}_{\mathsf{sr}}|^2\mathsf{SNR} + 1} < \frac{R}{\mathsf{SNR}}\} \qquad (23)\\
&\geq \mathbb{P}\{|h_{\mathsf{sd}}|^2 + |h_{\mathsf{rd}}|^2|h_{\mathsf{sr}}|^2\mathsf{SNR} < \frac{\mathsf{R}}{\mathsf{SNR}}\} \\
&\geq \mathbb{P}\left\{|h_{\mathsf{sd}}|^2 < \frac{R}{2\,\mathsf{SNR}}\right\} \times \mathbb{P}\left\{|h_{\mathsf{rd}}|^2 < \sqrt{\frac{R}{2(\mathsf{SNR})^2}}\right\} \times \mathbb{P}\left\{|h_{\mathsf{sr}}|^2 < \sqrt{\frac{R}{2(\mathsf{SNR})^2}}\right\}
\end{aligned}$$

(23) is true as $\ln(1+x) \leq x$. Also from (23) $\epsilon \to 0$ implies that $\frac{R}{\mathsf{SNR}} \to 0$, thus we have

$$\begin{aligned}
\lim_{\substack{\epsilon\to 0\\ SNR\to 0}} \frac{P_{out_{AF}}(R)}{\left(\frac{R}{\mathsf{SNR}}\right)^2} &\geq \lim_{\substack{\mathsf{SNR}\to 0\\ \frac{R}{\mathsf{SNR}}\to 0}} \frac{\left(1 - e^{-g_{\mathsf{sd}}^{-1}\frac{R}{2\,\mathsf{SNR}}}\right)\left(1 - e^{-g_{\mathsf{rd}}^{-1}\sqrt{\frac{R}{2\,\mathsf{SNR}^2}}}\right)\left(1 - e^{-g_{\mathsf{sr}}^{-1}\sqrt{\frac{R}{2\,\mathsf{SNR}^2}}}\right)}{\left(\frac{R}{\mathsf{SNR}}\right)^2} \\
&\approx \lim_{\mathsf{SNR}\to 0} \frac{g_{\mathsf{sd}}^{-1} g_{\mathsf{rd}}^{-1} g_{\mathsf{sr}}^{-1}}{4\mathsf{SNR}} \to \infty
\end{aligned}$$

This shows that

$$\lim_{\substack{\epsilon\to 0\\ \mathsf{SNR}\to 0}} \frac{R_{\epsilon_{AF}}}{\sqrt{\epsilon}\,\mathsf{SNR}} = 0 \tag{24}$$

□

Although amplify-forward strategy is able to provide full diversity at high SNR regime [3], this theorem clarifies that it does not provide full diversity at low SNR. The reason is that at low SNR regime, most of the received signal at the relay is noise. So relay becomes useless by transmitting mostly noise than signal. To improve the performance of this scheme one might think of transmitting bursty signals at the source to help the relay receive a less noisy observation. This scheme is called bursty amplify-forward (BAF) and is considered in next section.

## 4.4 The Achievable Outage Rate of Bursty Amplify-Forward Protocol

As it was mentioned before, the key point that makes AF protocol not suitable for low SNR is that the relay mostly injects noise to the system and it becomes useless. To overcome this fact we look at the scheme that the source does bursty transmission. Thus, the source broadcasts the message in only a fraction of the time, $\alpha$, with high power and it remains silent for the rest of the time. As the transmission power (while transmitting) is $\frac{P}{\alpha}$ the achievable rate (nats/s) using this strategy for a fixed realization of the channels is

$$I_{BAF}(\alpha) = \alpha \ln\left[1 + \frac{P}{\alpha N}\left(|h_{\mathsf{sd}}|^2 + \frac{|h_{\mathsf{rd}}|^2|h_{\mathsf{sr}}|^2 P}{(|h_{\mathsf{rd}}|^2 + |h_{\mathsf{sr}}|^2)P + \alpha N}\right)\right] \tag{25}$$

Now we investigate the achievable outage rate of this protocol.

**Theorem 4.4.** *We can choose $\alpha$ as a function of $\mathsf{SNR}$ such that the achievable $\epsilon -$ outage rate of Bursty amplify-forward (BAF) scheme, $R_{\epsilon_{BAF}}$, satisfies:*

$$\lim_{\substack{\mathsf{SNR}\to 0\\ \epsilon\to 0\\ \alpha\to 0}} \frac{R_{\epsilon_{BAF}}(\alpha)}{\sqrt{\epsilon}\,\mathsf{SNR}} = \sqrt{\frac{2g_{\mathsf{sd}}g_{\mathsf{rd}}g_{\mathsf{sr}}}{g_{\mathsf{rd}} + g_{\mathsf{sr}}}} \tag{26}$$

*Proof.* We know that $P_{out_{BAF}}(R,\alpha) = \mathbb{P}\{I_{BAF}(\alpha) < R\}$ therefore from (25) we have

$$\mathbb{P}\{I_{BAF}(\alpha) < R\} = \mathbb{P}\{|h_{\mathsf{sd}}|^2 + \frac{|h_{\mathsf{rd}}|^2|h_{\mathsf{sr}}|^2}{|h_{\mathsf{rd}}|^2 + |h_{\mathsf{sr}}|^2 + \frac{\alpha}{\mathsf{SNR}}} < (e^{\frac{R}{\alpha}} - 1)\frac{\alpha}{\mathsf{SNR}}\} \tag{27}$$

Now as $(e^{\frac{R}{\alpha}} - 1)\frac{\alpha}{\mathsf{SNR}} \geq \frac{R}{\mathsf{SNR}}$ thus $\epsilon \to 0$ implies that $\frac{R}{\mathsf{SNR}} \to 0$. We pick $\alpha \to 0$ in such a way that $\delta := \frac{\alpha}{\mathsf{SNR}} \to 0$ and $\frac{R}{\alpha} \to 0$ (for example $\alpha = \sqrt{R\,\mathsf{SNR}}$ has this condition). Now by applying Lemma A.3 with $g(\delta) = \delta(e^{\frac{R}{\alpha}} - 1)$ we get

$$\lim_{\substack{\mathsf{SNR}\to 0\\ \frac{R}{\mathsf{SNR}}\to 0\\ \alpha\to 0}} \frac{P_{out_{BAF}}(R,\alpha)}{g^2(\delta)} = \lim_{\substack{\mathsf{SNR}\to 0\\ \frac{R}{\mathsf{SNR}}\to 0\\ \alpha\to 0}} \frac{P_{out_{BAF}}(R,\alpha)}{(\frac{R}{\mathsf{SNR}})^2} = \frac{g_{\mathsf{rd}} + g_{\mathsf{sr}}}{2g_{\mathsf{sd}}g_{\mathsf{rd}}g_{\mathsf{rd}}} \tag{28}$$

So we have the desired result,

$$\lim_{\substack{\mathsf{SNR}\to 0\\ \epsilon\to 0\\ \alpha\to 0}} \frac{R_{\epsilon_{BAF}}(\alpha)}{\sqrt{\epsilon}\,\mathsf{SNR}} = \sqrt{\frac{2g_{\mathsf{sd}}g_{\mathsf{rd}}g_{\mathsf{sr}}}{g_{\mathsf{rd}} + g_{\mathsf{sr}}}} \tag{29}$$

□

This theorem shows that at low SNR, for small outage probabilities, $\epsilon$, the achievable outage rate of BAF is:

$$R_{BAF} \approx \sqrt{\frac{2\mathsf{SNR}_{\mathsf{sd}}\mathsf{SNR}_{\mathsf{rd}}\mathsf{SNR}_{\mathsf{sr}}}{\mathsf{SNR}_{\mathsf{rd}} + \mathsf{SNR}_{\mathsf{sr}}}\,\epsilon} \tag{30}$$

Thus the outage performance of BAF protocol matches the max-flow min-cut upper bound on the $\epsilon$-*outage capacity* of the FD fading relay channel therefore we have proved the main theorem (1.1) for the $\epsilon$ outage capacity of FD fading relay channel. It is interesting to note that in [7] it was also shown that low duty cycle transmission can improve the performance of the amplify-forward scheme at low SNR when we have no fading. But there the achievable rate of this protocol did not match the max-flow min-cut upper bound. The reason for optimality of this protocol in outage behavior is that by picking parameter $\alpha$ we make sure that the effective rate of transmission, $\frac{R}{\alpha}$, is still small but the transmit power is quite high. Hence as the effective rate of transmission is still small, in the outage event both the direct and indirect (source to relay to destination) path should have low overall gain. Therefore at lease one of the links in the indirect path and the direct path should be in deep fade. Thus the typical outage event is when the direct link and one of the links in the indirect path (source to relay or relay to destination) are at low SNR due to being in deep fade and the other link is at high SNR. In this case we can ignore the noise of the strong link in the indirect path. Also as the received SNR from both direct and indirect paths are low we are still energy efficient. Therefore the typical outage behavior of this protocol matches that of the cutset bound.

### 4.5 Outage Capacity per Unit Cost of FD Fading Relay Channel

We define the $\epsilon$-*outage* capacity per unit energy of the FD- relay channel to be the maximum number of bits that one can transmit with outage probability $\epsilon$, per unit energy spent at the source and unit energy spent at the relay. The previous results on the outage capacity directly apply to the outage capacity per unit cost of FD relay channel with fading, $\mathbf{C}_{\epsilon,\,relay}$, for small probability of outage. In FD relay channel, for any SNR, the maximum rate to have outage probability less than $\epsilon$ is $C_{\epsilon_{relay}}(\mathsf{SNR})$ which is obviously concave and strictly increasing in SNR (otherwise by time sharing we will get better rate). Thus

$$\mathbf{C}_{\epsilon,\,relay} = \sup_{\mathsf{SNR}} \frac{C_{\epsilon_{relay}}(\mathsf{SNR})}{\mathsf{SNR}}$$

is achieved by letting $\mathsf{SNR} \to 0$. Therefore

$$\mathbf{C}_{\epsilon,\,relay} = \lim_{\mathsf{SNR}\to 0} \frac{C_{\epsilon_{relay}}(\mathsf{SNR})}{\mathsf{SNR}}$$

Thus from our result on the $C_{\epsilon_{relay}}$, (1), we complete the proof of the Main Theorem

$$\lim_{\epsilon\to 0} \frac{\mathbf{C}_{\epsilon,\,relay}}{\sqrt{\epsilon}} = \lim_{\substack{\epsilon\to 0\\ \mathsf{SNR}\to 0}} \frac{C_{\epsilon_{relay}}}{\sqrt{\epsilon}\,\mathsf{SNR}} = \sqrt{\frac{2g_{\mathsf{sd}}g_{\mathsf{rd}}g_{\mathsf{sr}}}{g_{\mathsf{rd}}+g_{\mathsf{sr}}}}$$

## 5 The Effect of Channel Knowledge

Channel estimation is quite challenging in the low SNR regime therefore it is important to understand how much the channel knowledge is beneficial or crucial to the outage capacity of the fading relay channel in the interested regime. We study two extremes in this section. One extreme is the case that neither the transmitter nor the receiver knows the channel (non coherent model). We show even without the channel knowledge available at the destination one can achieve the same outage capacity as before using a bursty pulse position modulation (PPM) scheme. On the other hand in the other extreme that both the transmitter and receiver know the channel (Full CSI model) we show that the outage capacity can be increased slightly while the channel estimation becomes very hard. We also discuss that the bursty amplify-forward scheme combined with beam-forming can achieve the outage capacity in this case.

## 5.1 Outage Capacity of Non-Coherent Fading Relay Channel

In this section we show that even without the channel knowledge available at the receiver as well as the transmitter, we can achieve the same outage capacity as the case that CSI is available at the receiver (1). The achievable scheme is using bursty pulse position modulation coding and the amplify forward scheme at the relay. The detection at the destination is based on energy detection, i.e. the position with highest energy is decoded at the destination.

Let $\phi_1, \ldots, \phi_M$ be $M$ orthonormal signals of the form $\phi_i = (0, \ldots, 0, 1, 0, \ldots, 0)$, which is a length $M$ vector with non-zero value at the i-th position ($i = 1, \ldots, M$).

To transmit message $m$ ($m = 1, \ldots, M$), the source will broadcast the message $x_m = A\phi_m$ followed by zeros in $L > M$ time slots. The relay and the destination will respectively receive $y_R$ and $y_1$. In the next $L$ time slots, the source remains silent and the relay will transmit the first $M$ time unit of $y_R$ that contains information (normalized by $\sqrt{\frac{A^2}{A^2|h_{\mathsf{sr}}|^2+M}}$ to satisfy the average power and remains silent afterwards and the destination will receive $y_2$. In order to satisfy the average power constraint we should have $A^2 = 2LP$. To decode, the destination will compute $y = (|y_{1,1}|^2 + |y_{2,1}|^2/\hat{\sigma}^2, \ldots, |y_{1,M}|^2 + |y_{2,M}|^2/\hat{\sigma}^2)$, where $\hat{\sigma}^2$ is the estimated variance of the indirect path and $y_{1,m}$ and $y_{2,m}$ are respectively the projection of the first M elements of $y_1$ and $y_2$ onto $\phi_m$ ($m = 1, \ldots, M$) :

$$y_{1,m} = y_1(1, \ldots, M)\phi_m^t, \quad m = 1, \ldots, M \tag{31}$$

$$y_{2,m} = y_2(1, \ldots, M)\phi_m^t, \quad m = 1, \ldots, M \tag{32}$$

The destination will decode the unique message $m$ if the $m - th$ component of $y$ is maximum. But for making the analysis simpler we will use another decoding technique that requires the destination to pick a threshold, $\tau$, and to decode the unique message $m$ if the $m-th$ component of $y$ is uniquely larger than the threshold $\tau$. It is obvious that the probability of error using this genie aided scheme can not be less than the first strategy (picking the maximum).

By symmetry lets assume that message 1 has been transmitted by the source, then for fixed channel gains we have,

$$\begin{aligned}
y_{1,1} &= Ah_{\mathsf{sd}} + z_{1,1} \\
&\sim \mathcal{CN}(Ah_{\mathsf{sd}}, 1) \\
y_{1,m} &= z_{1,m}, \quad 1 < m \leq M \\
&\sim \mathcal{CN}(0, 1) \\
y_{2,1} &= \frac{A^2 h_{\mathsf{sr}} h_{\mathsf{rd}}}{\sqrt{A^2|h_{\mathsf{sr}}|^2 + M}} + \frac{Ah_{\mathsf{rd}}}{\sqrt{A^2|h_{\mathsf{sr}}|^2 + M}} z_{R,1} + z_{2,1} \\
&\sim \mathcal{CN}(\frac{A^2 h_{\mathsf{sr}} h_{\mathsf{rd}}}{\sqrt{A^2|h_{\mathsf{sr}}|^2 + M}}, \frac{A^2|h_{\mathsf{rd}}|^2}{A^2|h_{\mathsf{sr}}|^2 + M} + 1) \\
y_{2,m} &= \frac{Ah_{\mathsf{rd}}}{\sqrt{A^2|h_{\mathsf{sr}}|^2 + M}} z_{R,m} + z_{2,m}, \quad 1 < m \leq M \\
&\sim \mathcal{CN}(0, \frac{A^2|h_{\mathsf{rd}}|^2}{A^2|h_{\mathsf{sr}}|^2 + M} + 1)
\end{aligned}$$

where $z_{1,1}, z_{1,2}, z_{2,1}, z_{2,2}, z_{R,1}$ and $z_{R,2}$ are distributed like $\mathcal{CN}(0, 1)$.

There are two cases that the decoding fails:

$$|y_{1,1}|^2 + \frac{|y_{2,1}|^2}{\hat{\sigma}^2} < \tau \tag{33}$$

or there exists one $1 < i \leq M$ such that

$$|y_{1,i}|^2 + \frac{|y_{2,i}|^2}{\hat{\sigma}^2} > \tau \tag{34}$$

We select the variance estimator to be

$$\hat{\sigma}^2 = \frac{\sum_{i=1}^{M} |y_{2,i}|^2}{M} \overset{\text{M Large}}{\approx} \frac{A^2|h_{\mathsf{rd}}|^2}{A^2|h_{\mathsf{sr}}|^2 + M} (\frac{A^2|h_{\mathsf{sr}}|^2}{M} + 1) + 1 \tag{35}$$

Now to make the probability of first event small we make sure that the mean of the random variable $|y_{1,1}|^2 + |y_{2,1}|^2/\hat{\sigma}^2$ is far from $\tau$.

$$\begin{aligned}
\mathbb{E}[|y_{1,1}|^2 + \frac{|y_{2,1}|^2}{\hat{\sigma}^2}] &\approx A^2|h_{\mathsf{sd}}|^2 + \frac{A^4|h_{\mathsf{sr}}|^2|h_{\mathsf{rd}}|^2}{A^2|h_{\mathsf{sr}}|^2 + A^2|h_{\mathsf{rd}}|^2 + M + \frac{A^4|h_{\mathsf{sr}}|^2|h_{\mathsf{rd}}|^2}{M}} + \\
&\quad + \frac{A^2|h_{\mathsf{sr}}|^2 + A^2|h_{\mathsf{rd}}|^2 + M}{A^2|h_{\mathsf{sr}}|^2 + A^2|h_{\mathsf{rd}}|^2 + M + \frac{A^4|h_{\mathsf{sr}}|^2|h_{\mathsf{rd}}|^2}{M}} \\
&> A^2|h_{\mathsf{sd}}|^2 + \frac{A^4|h_{\mathsf{sr}}|^2|h_{\mathsf{rd}}|^2}{A^2|h_{\mathsf{sr}}|^2 + A^2|h_{\mathsf{rd}}|^2 + M + \frac{A^4|h_{\mathsf{sr}}|^2|h_{\mathsf{rd}}|^2}{M}}
\end{aligned}$$

To make the mean of the random variable $|y_{1,1}|^2 + \frac{|y_{2,1}|^2}{\hat{\sigma}^2}$ far from $\tau$ it is sufficient to have

$$\tau << A^2|h_{\mathsf{sd}}|^2 + \frac{A^4|h_{\mathsf{sr}}|^2|h_{\mathsf{rd}}|^2}{A^2|h_{\mathsf{sr}}|^2 + A^2|h_{\mathsf{rd}}|^2 + M + \frac{A^4|h_{\mathsf{sr}}|^2|h_{\mathsf{rd}}|^2}{M}} \quad (36)$$

Before considering the second case we state a lemma,

**Lemma 5.1.** *If $u$ and $v$ are exponential random variables with mean $\mu_u$ and $\mu_v$ respectively then*

$$\mathbb{P}\{u + v > \tau\} = \frac{\mu_u}{\mu_u - \mu_v} e^{-\frac{\tau}{\mu_u}} - \frac{\mu_v}{\mu_u - \mu_v} e^{-\frac{\tau}{\mu_v}} \quad (37)$$

*Proof.*

$$\begin{aligned}
\mathbb{P}\{u + v < \tau\} &= \int_0^\tau \frac{1}{\mu_u} e^{-\frac{x}{\mu_u}} \mathbb{P}\{v < \tau - x\} dx \\
&= \int_0^\tau \frac{1}{\mu_u} e^{-\frac{x}{\mu_u}} (1 - e^{-\frac{\tau - x}{\mu_v}}) dx \\
&= 1 - e^{\frac{\tau}{\mu_u}} - \frac{1}{\mu_u} e^{-\frac{\tau}{\mu_v}} \int_0^\tau e^{(\frac{1}{\mu_v} - \frac{1}{\mu_u})x} dx \\
&= 1 - \frac{\mu_u}{\mu_u - \mu_v} e^{-\frac{\tau}{\mu_u}} + \frac{\mu_v}{\mu_u - \mu_v} e^{-\frac{\tau}{\mu_v}}
\end{aligned}$$

□

The second case of error consists of $M - 1$ events. Each event occurs when a sum of two exponential random variables (with means 1 and $\mu = \frac{\frac{A^2|h_{\mathsf{rd}}|^2}{A^2|h_{\mathsf{sr}}|^2 + M} + 1}{\hat{\sigma}^2} < 1$) is greater than $\tau$. Therefore by union bound and lemma 5.1 we have

$$
\begin{aligned}
\mathbb{P}\{\exists i : |y_{1,i}|^2 + \frac{|y_{2,i}|^2}{\hat{\sigma}^2} > \tau\} &< (M-1)\mathbb{P}\{|y_{1,2}|^2 + \frac{|y_{2,2}|^2}{\hat{\sigma}^2} > \tau\} \\
&< \frac{1}{1-\mu} e^{\ln M - \tau}
\end{aligned}
$$

Now in order to make the probability of this event small we should have a large negative exponent, i.e.

$$
\ln M << \tau \tag{38}
$$

To be able to pick the threshold $\tau$ to simultaneously satisfy (36) and (38) we should have

$$
\ln M << A^2|h_{\mathsf{sd}}|^2 + \frac{A^4|h_{\mathsf{sr}}|^2|h_{\mathsf{rd}}|^2}{A^2|h_{\mathsf{sr}}|^2 + A^2|h_{\mathsf{rd}}|^2 + M + \frac{A^4|h_{\mathsf{sr}}|^2|h_{\mathsf{rd}}|^2}{M}} \tag{39}
$$

As we are interested in the regime that $R = \frac{\ln M}{2L} \to 0$, $P \to 0$ and $\frac{R}{P} = \frac{\log M}{2LP} \to 0$, we pick $M$ and $L$ large enough such that:

$$
\ln M << M << 2LP = A^2 \tag{40}
$$

Therefore long as $\min\{A^2|h_{\mathsf{sr}}|^2, A^2|h_{\mathsf{rd}}|^2\} < M$,

$$
\frac{A^4|h_{\mathsf{sr}}|^2|h_{\mathsf{rd}}|^2}{A^2|h_{\mathsf{sr}}|^2 + A^2|h_{\mathsf{rd}}|^2 + M + \frac{A^4|h_{\mathsf{sr}}|^2|h_{\mathsf{rd}}|^2}{M}} \approx \min\{A^2|h_{\mathsf{sr}}|^2, A^2|h_{\mathsf{rd}}|^2\} \tag{41}
$$

and we can satisfy (39) if

$$
\ln M << A^2|h_{\mathsf{sd}}|^2 + A^2 \min\{|h_{\mathsf{sr}}|^2, |h_{\mathsf{rd}}|^2\} \tag{42}
$$

or

$$
\frac{R}{P} = \frac{\ln M}{2LP} = \frac{\ln M}{A^2} < |h_{\mathsf{sd}}|^2 + \min\{|h_{\mathsf{sr}}|^2, |h_{\mathsf{rd}}|^2\} \tag{43}
$$

which is the same expression as the max-flow min-cut upper bound (11) being greater than the rate that we try to communicate.

And if $\min\{A^2|h_{\mathsf{sr}}|^2, A^2|h_{\mathsf{rd}}|^2\} > M$ then

$$
\frac{A^4|h_{\mathsf{sr}}|^2|h_{\mathsf{rd}}|^2}{A^2|h_{\mathsf{sr}}|^2 + A^2|h_{\mathsf{rd}}|^2 + M + \frac{A^4|h_{\mathsf{sr}}|^2|h_{\mathsf{rd}}|^2}{M}} \approx M \tag{44}
$$

and we can obviously satisfy (39).

## 5.2 Outage Capacity of the Fading Relay Channel with Full CSI

In this part we investigate the outage capacity of the FD-relay channel shown in Figure 1. The model is the same as before except for the fact that channel state information is available at both transmitter and receiver (full CSI). One might think once the transmitter knows the channel it can always adjust the power such that no outage occurs, this is a valid idea if we can average the power on different realizations of the channel. How ever in a slow fading scenario that the channel varies very slowly over time it is a practical constraint to have average power constraint during a single realization of the channel. Therefore in this scenario the transmitter can not avoid the outage and outage capacity if an interesting measure to look at.

### 5.2.1 The Upper Bound on the Outage Capacity

In this section we use the general max-flow min-cut bound for the network shown in Figure 1 to find an upper bound on the $\epsilon - outage$ capacity of the FD relay channel with full CSI using in the limit of low SNR and low probability of outage. For fixed channel gains, the max-flow min-cut bound is as follows:

$$C_{realy}(h_{\mathsf{sr}}, h_{\mathsf{sd}}, h_{\mathsf{rd}}) \leq \max_{p(X_1, X_2, X_R)} \min(I(X_1, X_2, X_R; Y_1, Y_2), I(X_1, X_2; Y_R, Y_1, Y_2|X_R)) \quad (45)$$

The first term which is corresponding to the multiple access is bounded by:

$$\begin{aligned}
I(X_1, X_2, X_R; Y_1, Y_2) &= h(Y_1, Y_2) - h(Y_1, Y_2|X_1, X_2, X_R) \\
&= h(Y_1, Y_2) - h(Z_1, Z_2) \\
&= h(Y_1, Y_2) - \ln(2\pi e) \\
&\leq \alpha \ln\left(var(Y_1)\right) + (1-\alpha)\ln\left(var(Y_2)\right) \\
&\leq \alpha \ln\left(1 + |h_{\mathsf{sd}}|^2 var(X_1)\right) + \\
&\quad +(1-\alpha)\ln\left(1 + |h_{\mathsf{sd}}|^2 var(X_2) + |h_{\mathsf{rd}}|^2 var(X_R) + 2|h_{\mathsf{sd}}||h_{\mathsf{rd}}|\rho E(X_2 X_R)\right) \\
&\leq \alpha \ln\left(1 + |h_{\mathsf{sd}}|^2 P_1\right) + \\
&\quad +(1-\alpha)\ln\left(1 + |h_{\mathsf{sd}}|^2 P_2 + |h_{\mathsf{rd}}|^2 P_R + 2|h_{\mathsf{sd}}||h_{\mathsf{rd}}|\rho\sqrt{P_2 P_R}\right) \\
&\leq \alpha |h_{\mathsf{sd}}|^2 P_1 + (1-\alpha)(|h_{\mathsf{sd}}|^2 P_2 + |h_{\mathsf{rd}}|^2 P_R + 2|h_{\mathsf{sd}}||h_{\mathsf{rd}}|\rho\sqrt{P_2 P_R}) \\
&= \beta |h_{\mathsf{sd}}|^2 P + ((1-\beta)|h_{\mathsf{sd}}|^2 P + |h_{\mathsf{rd}}|^2 P + 2|h_{\mathsf{sd}}||h_{\mathsf{rd}}|\rho\sqrt{1-\beta})P \qquad (46)
\end{aligned}$$

where we define

$$\rho = \frac{E(X_2 X_R)}{\sqrt{E(X_2^2)E(X_R^2)}} \qquad (47)$$

Now we will bound the second term which is corresponding to the broadcast cut. First we note that given $X_R$ we have two parallel channels: $(X_1; Y_1, Y_R)$ and $(X_2; Y_2)$ therefore we have the following markov chain: $Y_2 \to X_2 \to X_1 \to (Y_1, Y_R)$ which makes the following equalities obvious:

$$I(X_2; (Y_1, Y_R)|X_1, X_R) = 0$$

$$0 \leq I(X_1; Y_2|(Y_1, Y_R), X_2, X_R) \leq I(X_1, (Y_1, Y_R); Y_2|X_2, X_R) = 0$$

Therefore,

$$\begin{aligned} I(X_1, X_2; Y_R, Y_1, Y_2|X_R) &= I(X_1; (Y_1, Y_R)|X_R) + I(X_2; (Y_1, Y_R)|X_1, X_R) \\ &\quad + I(X_2; Y_2|(Y_1, Y_R), X_R) + I(X_1; Y_2|(Y_1, Y_R), X_2, X_R) \quad (48) \\ &= I(X_1; (Y_1, Y_R)|X_R) + I(X_2; Y_2|(Y_1, Y_R), X_R) \quad (49) \\ &= I(X_1; (Y_1, Y_R)|X_R) + I(X_2, (Y_1, Y_R); Y_2|X_R) - I((Y_1, Y_R); Y_2|X_R) \\ &= I(X_1; (Y_1, Y_R)|X_R) + I(X_2; Y_2|X_R) - I((Y_1, Y_R); Y_2|X_R) \quad (50) \\ &\leq I(X_1; (Y_1, Y_R)|X_R) + I(X_2; Y_2|X_R) \quad (51) \end{aligned}$$

Now we have:

$$I(X_1; Y_1, Y_R|X_R) \leq \alpha \ln(1 + (|h_{\mathsf{sd}}|^2 + |h_{\mathsf{sr}}|^2)P_1)$$

and

$$\begin{aligned} I(X_2; Y_2|X_R) &\leq (1-\alpha)\ln(1 + |h_{\mathsf{sd}}|^2(E(X_2^2) - \frac{(E(X_2 X_R))^2}{E(X_R^2)}) \\ &= (1-\alpha)\ln(1 + |h_{\mathsf{sd}}|^2(1-\rho^2)P_2) \\ &\leq (1-\alpha)|h_{\mathsf{sd}}|^2(1-\rho^2)P_2 \\ &= (1-\beta)|h_{\mathsf{sd}}|^2(1-\rho^2)P \quad (52) \end{aligned}$$

Therefore,

$$\begin{aligned} C_{relay}(h_{\mathsf{sd}}, h_{\mathsf{rd}}, h_{\mathsf{sr}}) \leq \min_{\substack{0\leq\beta\leq 1 \\ -1\leq\rho\leq 1}} \{&|h_{\mathsf{sd}}|^2(\beta + (1-\beta)(1-\rho^2)) + |h_{\mathsf{sr}}|^2\beta, |h_{\mathsf{sd}}|^2 + |h_{\mathsf{rd}}|^2 + \\ &+ 2|h_{\mathsf{sd}}||h_{\mathsf{rd}}|\rho\sqrt{1-\beta}\}P \end{aligned} \quad (53)$$

Now we have the following bound on the outage probability with full CSI:

$$\begin{aligned} P_{out,relay} \geq \min_{\substack{0\leq\beta\leq 1 \\ -1\leq\rho\leq 1}} \mathbb{P}\{&|h_{\mathsf{sd}}|^2(\beta + (1-\beta)(1-\rho^2)) + |h_{\mathsf{sr}}|^2\beta, |h_{\mathsf{sd}}|^2 + |h_{\mathsf{rd}}|^2 + \\ &+ 2|h_{\mathsf{sd}}||h_{\mathsf{rd}}|\rho\sqrt{1-\beta} < \frac{R}{SNR}\} \end{aligned} \quad (54)$$

### 5.2.2 The Achievable Scheme: BAF + Beamforming

Here we show that for any choice of $\beta$ and $\rho$ it is possible to achieve the max-flow min-cut bound on the outage probability shown in (54) in the limit of low SNR and low outage probability. To achieve the bound we use the described BAF protocol (source talks fraction of $\alpha$ of the time) with the difference that here the source uses both frequency bands to transmit the new data and in the second frequency band some of the power is allocated to beamform with the help of the relay.

For given $\beta$ and $\rho$, we construct $X_1$ using random Gaussian code generation with power $\frac{\beta P}{\alpha}$. Now a part of $X_2$ should be used to transmit new data and a part of it is used to beamform with the relay. Therefore we set,

$$X_2 = c_1 \frac{h_{\mathsf{sr}} h_{\mathsf{rd}}}{h_{\mathsf{sd}}} X_1 + \hat{X}_2 \tag{55}$$

where $\hat{X}_2$ is another codeword generated using random Gaussian codeword generation with power $\frac{(1-\rho^2)(1-\beta)P}{\alpha}$ and $c_1$ is a constant to set the power constraint on $X_2$ equal to $\frac{(1-\beta)P}{\alpha}$. And as we are using Amplify-Forward protocol

$$X_R = \frac{Y_R}{\sqrt{|g|^2 \frac{\beta P}{\alpha} + 1}} \sqrt{\frac{P}{\alpha}} \tag{56}$$

Therefore we have,

$$\begin{aligned} Y_R &= h_{\mathsf{sr}} X_1 + Z_R \\ Y_1 &= h_{\mathsf{sd}} X_1 + Z_1 \\ Y_2 &= h_{\mathsf{sd}} X_2 + h_{\mathsf{rd}} X_R + Z_2 \\ &= c_1 h_{\mathsf{sr}} h_{\mathsf{rd}} X_1 + h_{\mathsf{sd}} \hat{X}_2 + \frac{h_{\mathsf{sr}} h_{\mathsf{rd}}}{\sqrt{|h_{\mathsf{sr}}|^2 \frac{\beta P}{\alpha} + 1}} \sqrt{\frac{P}{\alpha}} X_1 + \frac{h_{\mathsf{rd}}}{\sqrt{|h_{\mathsf{sr}}|^2 \frac{\beta P}{\alpha} + 1}} \sqrt{\frac{P}{\alpha}} Z_R + Z_2 \end{aligned}$$

Now if we write the equations in the matrix form,

$$\begin{pmatrix} Y_1 \\ Y_2 \end{pmatrix} = \begin{pmatrix} h_{\mathsf{sd}} & 0 \\ h_{\mathsf{sr}} h_{\mathsf{rd}} \left(c_1 + \sqrt{\frac{P}{|h_{\mathsf{sr}}|^2 \beta P + \alpha}}\right) & h_{\mathsf{sd}} \end{pmatrix} \begin{pmatrix} X_1 \\ \hat{X}_2 \end{pmatrix} + \begin{pmatrix} Z_1 \\ h_{\mathsf{rd}} \sqrt{\frac{P}{|h_{\mathsf{sr}}|^2 \beta P + \alpha}} Z_R + Z_2 \end{pmatrix} \tag{57}$$

and

$$c_1^2 = \frac{|h_{\mathsf{sd}}|^2(1-\beta)\rho^2}{\beta|h_{\mathsf{sr}}|^2|h_{\mathsf{rd}}|^2} \tag{58}$$

The capacity of this $2 \times 2$ MIMO system is :

$$R_{\text{BAF+B}} = \alpha \ln(1 + |h_{\mathsf{sd}}|^2\frac{\beta P}{\alpha} + \frac{|h_{\mathsf{sd}}|^2(|h_{\mathsf{sr}}|^2\beta P+\alpha)((1-\rho^2)(1-\beta)P))}{\alpha(|h_{\mathsf{sr}}|^2\beta P + |h_{\mathsf{rd}}|^2 P+\alpha)} + \\ + \frac{|h_{\mathsf{sr}}|^2|h_{\mathsf{rd}}|^2(c_1\sqrt{|h_{\mathsf{sr}}|^2\beta P+\alpha}+\sqrt{P})^2(\beta P)}{\alpha(|h_{\mathsf{sr}}|^2\beta P + |h_{\mathsf{rd}}|^2 P+\alpha)} + |h_{\mathsf{sd}}|^2\frac{\beta P}{\alpha}\frac{|h_{\mathsf{sd}}|^2(|h_{\mathsf{sr}}|^2\beta P+\alpha)((1-\rho^2)(1-\beta)P)}{\alpha(|h_{\mathsf{sr}}|^2\beta P + |h_{\mathsf{rd}}|^2 P+\alpha)}) \tag{59}$$

Now we can verify that in the cases of typical outage this achievable rate has the same shape as the max-flow min-cut bound. For example one case of typical outage is when both $h_{\mathsf{sd}}$ and $h_{\mathsf{rd}}$ are weak and $h_{\mathsf{sr}}$ is not weak. In this case we have,

$$\begin{aligned} R_{\text{BAF+B}} &\approx \alpha \ln(1 + |h_{\mathsf{sd}}|^2\frac{\beta P}{\alpha} + |h_{\mathsf{sd}}|^2(1-\rho)^2(1-\beta)\frac{P}{\alpha} + |h_{\mathsf{sd}}|^2(1-\beta)\rho^2\frac{P}{\alpha} + \\ &\quad + |h_{\mathsf{rd}}|^2\frac{P}{\alpha} + 2|h_{\mathsf{sd}}||h_{\mathsf{rd}}|\rho\sqrt{1-\beta}\frac{P}{\alpha}) \\ &\approx |h_{\mathsf{sd}}|^2 P + |h_{\mathsf{rd}}|^2 P + 2|h_{\mathsf{sd}}||h_{\mathsf{rd}}|\rho\sqrt{1-\beta}P \end{aligned} \tag{60}$$

Which is the same as the multiple access cut in (53).

To have a sense of this additional gain lets look at the case that all the channel gains are rayleigh fading with variance 1 ($g_{\mathsf{sd}} = g_{\mathsf{sr}} = g_{\mathsf{rd}} = 1$). If we solve the maximization problem in this case we get

$$\beta \approx 0.94$$

$$\rho \approx 1$$

$$R_{\text{BAF+B}} \approx 1.04\sqrt{\epsilon}\,\mathsf{SNR}$$

Now if we compared this rate to the outage capacity of the corresponding relay channel without CSI at the transmitter ($\approx \sqrt{\epsilon}\,\mathsf{SNR}$) we notice that the additional gain from having CSI at the transmitter is quite low (just 4%). This can be intuitively explained by noticing that the source prefers to allocate more power to the first frequency band which both the destination and the relay can receive data from to increase diversity than the second frequency band (for beam-forming with the relay).

# 6 Extensions

In this section we look at two important extensions: First we look at the scenario that we have average power constraint on the source and the relay and the question is what is the optimal power allocation to them in the sense of outage capacity. As the second extension we look at the network where the transmission from the source to the destination is helped by $k$ relays. To understand the gain obtained by adding more relays into the network, we compute the outage capacity of this network in the interested regime. We also show that the same BAF protocol is optimal and achieves the outage capacity of this network in the interested regime.

## 6.1 Optimized Power Allocation to the Source and the Relay

Sometimes in the case of limited energy scenarios, there is a sum energy constraint on the transmitting nodes and it is important to optimize the energy spent at each transmitting node. In the case of the relay channel we consider the case that we have the average some power constraint equal to $2P$ on the source and the relay. But from the behavior of Bursty Amplify-Forward protocol we know that for each power allocation the max-flow min-cut bound corresponding to that power allocation is tight in the limit of low SNR and low outage probability. Therefore it just remains to minimize the outage probability corresponding to rate equal to max-flow min-cut bound and the achievability of this outage probability is guaranteed by BAF protocol.

If power used for transmission at the source and the relay are represented by $P_1$ and $P_2$ respectively then to satisfy the average sum power constraint we have $\frac{1}{2}(P_1 + P_2) = 2P$. The max-flow min-cut bound is:

$$\begin{aligned} C_{relay}(h_{\mathsf{sd}}, h_{\mathsf{rd}}, h_{\mathsf{sr}}) &\leq \min(\frac{1}{2}\ln(1 + (|h_{\mathsf{sd}}|^2 + |h_{\mathsf{sr}}|^2)P_1), \frac{1}{2}\ln(1 + |h_{\mathsf{sd}}|^2 P_1) + \frac{1}{2}\ln(1 + |h_{\mathsf{rd}}|^2 P_2)) \\ &\leq \frac{1}{2}(|h_{\mathsf{sd}}|^2 P_1 + \min(|h_{\mathsf{sr}}|^2 P_1, |h_{\mathsf{rd}}|^2 P_2)) \\ &= \left(|h_{\mathsf{sd}}|^2 \beta + \min(|h_{\mathsf{sr}}|^2 \beta, |h_{\mathsf{rd}}|^2 (1-\beta))\right) 2P \end{aligned} \tag{61}$$

where $\frac{1}{2}P_1 = \beta 2P$ and $\frac{1}{2}P_2 = (1-\beta)2P$. Therefore,

$$\begin{aligned}
P_{out_{relay}}(R) &= \mathbb{P}\{C_{relay}(h_{\mathsf{sr}}, h_{\mathsf{sd}}, h_{\mathsf{rd}}) < R\} \\
&\geq \mathbb{P}\{2\mathsf{SNR}(|\mathsf{h}_{\mathsf{sd}}|^2\beta + \min(|\mathsf{h}_{\mathsf{sr}}|^2\beta, |\mathsf{h}_{\mathsf{rd}}|^2(1-\beta))) < \mathsf{R}\} \\
&= \mathbb{P}\{|h_{\mathsf{sd}}|^2\beta + \min(|h_{\mathsf{sr}}|^2\beta, |h_{\mathsf{rd}}|^2(1-\beta)) < \frac{R}{2\mathsf{SNR}}\}
\end{aligned}$$

Now $\min(|h_{\mathsf{sr}}|^2\beta, |h_{\mathsf{rd}}|^2(1-\beta))$ is another exponential random variable with mean $\frac{\beta(1-\beta)g_{\mathsf{rd}}g_{\mathsf{rd}}}{\beta g_{\mathsf{sr}}+(1-\beta)g_{\mathsf{rd}}}$. Using Lemma A.1 we have

$$\lim_{\substack{\mathsf{SNR}\to 0 \\ \frac{R}{\mathsf{SNR}}\to 0}} \frac{p_{out_{relay}}(R)}{\left(\frac{R}{\mathsf{SNR}}\right)^2} \geq \frac{1}{4 \times 2 \times \beta g_{\mathsf{sd}} \times \frac{\beta(1-\beta)g_{\mathsf{rd}}g_{\mathsf{sr}}}{\beta g_{\mathsf{sr}}+(1-\beta)g_{\mathsf{rd}}}} = \frac{\beta g_{\mathsf{sr}} + (1-\beta)g_{\mathsf{rd}}}{8\beta^2(1-\beta)g_{\mathsf{sd}}g_{\mathsf{rd}}g_{\mathsf{sr}}} \tag{62}$$

In order to minimize the outage probability, we should minimize the term shown in (62) with respect to $0 \leq \beta \leq 1$, which is very simple.

## 6.2 Network With k Relays

In this section we look at a network consisting of a source (S) and a destination (D) and $k$ relays $(R_1, \ldots R_k)$ with half-duplex constraint. As we are interested in the low SNR regime, with the same intuition as before we can argue that without being sup optimal we can assume that all the communicating channels to the destination are orthogonal in frequency and each one has access to $\frac{1}{k+1}$ of the total bandwidth. Therefore the equivalent model is shown in figure (4). The channel gains $h_{\mathsf{sr_1}}, \ldots, h_{\mathsf{sr_k}}$ and $h_{\mathsf{sd}}, h_{\mathsf{r_1 D}}, \ldots, h_{\mathsf{r_k d}}$ are subject to independent Rayleigh fading with variances $g_{\mathsf{sr_1}}, \ldots, g_{\mathsf{sr_k}}$ and $g_{\mathsf{sd}}, g_{\mathsf{r_1 D}}, \ldots, g_{\mathsf{r_k d}}$ respectively.

### 6.2.1 Upper bound on the outage capacity

Any cut from the source to the destination will include the direct path $(h_{\mathsf{sd}})$ and exactly one of the channels in each indirect path $(h_{\mathsf{sr_i}}$ or $h_{\mathsf{r_i d}})$. Therefore there are $2^k$ different total cuts and the minimum of them will be the max-flow min-cut bound for this network. It can be easily shown by induction that for fixed channel gains, the max-flow min-cut upper bound

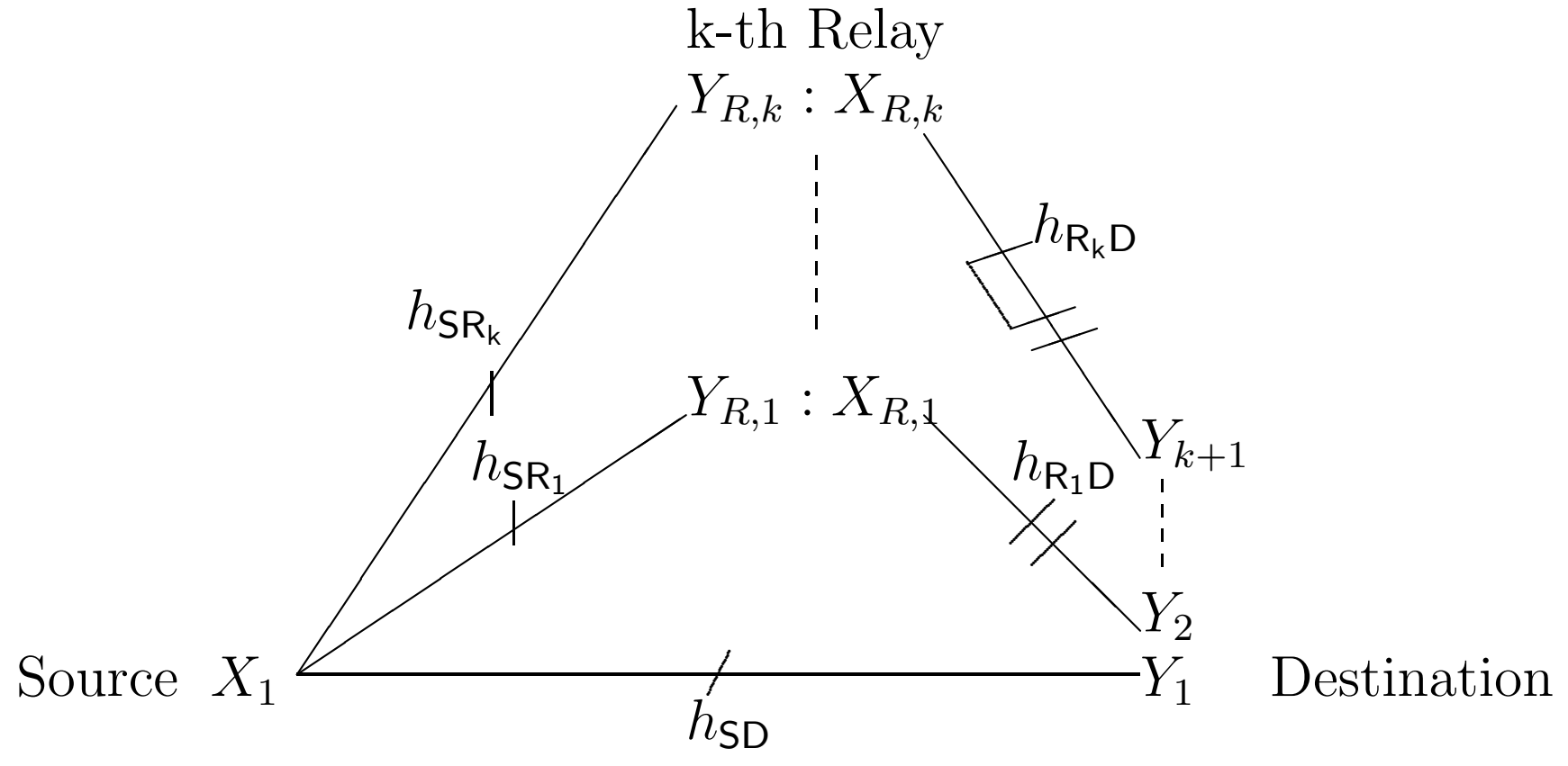

Figure 4: The general communication model at low SNR with $k$ relays.

on the achievable rate of this network is:

$$\begin{aligned} C_{k-relay}(h_{\mathsf{sr_1}}, \ldots, h_{\mathsf{sr_k}}, h_{\mathsf{sd}}, h_{\mathsf{r_1D}}, \ldots, h_{\mathsf{r_kd}}) &\leq \frac{1}{k+1} \min\{\{\ln(1 + (|h_{\mathsf{sd}}|^2 + \sum_{i \in V} |h_{\mathsf{sr_i}}|^2)\ (k+1)P) \\ &\quad + \sum_{i \in (\{1,\ldots,k\}-V)} \ln(1 + |h_{\mathsf{r_id}}|^2 (k+1)P) | V \subseteq \{1, \ldots, k\}\}\} \\ &\leq (|h_{\mathsf{sd}}|^2 + \sum_{i=1}^{k} \min\{|h_{\mathsf{sr_i}}|^2, |h_{\mathsf{r_id}}|^2\})\mathsf{SNR} \end{aligned} \tag{63}$$

Using this bound we can find the corresponding upper bound on the outage capacity of the FD network with $k$ relays. But before that we need a few lemmas.

**Lemma 6.1.** *If $u$ and $v$ are two independent random variables such that*

$$\begin{aligned} \lim_{\epsilon \to 0} \frac{\mathbb{P}\{u < \epsilon\}}{\epsilon} &= \alpha_1 \\ \lim_{\epsilon \to 0} \frac{\mathbb{P}\{v < \epsilon\}}{\epsilon^k} &= \alpha_2 \end{aligned}$$

*then*

$$\lim_{\epsilon \to 0} \frac{\mathbb{P}\{u + v < \epsilon\}}{\epsilon^{k+1}} = \frac{\alpha_1 \alpha_2}{k+1}$$

**Lemma 6.2.** *If $u_1, u_2, \ldots, u_k$ are $k$ independent exponential random variables with means $g_{\mathsf{sd}}, \mu_2, \ldots, \mu_k$ respectively then*

$$\lim_{\epsilon \to 0} \frac{\mathbb{P}\{u_1 + u_2 + \ldots + u_k < \epsilon\}}{\epsilon^k} = \frac{1}{k!\ \mu_1 \mu_2 \ldots \mu_k}$$

*Proof.* Proof by induction on $k$ and using lemma 6.1. □

**Theorem 6.3.** *The* $\epsilon$-outage *capacity ,* $C_{\epsilon_{k-relay}}$*, of the FD- network with k relays (in nats/s) satisfies*

$$\lim_{\substack{\epsilon\to 0\\ \mathsf{SNR}\to 0}} \frac{C_{\epsilon_{k-relay}}}{\sqrt[k+1]{\epsilon}\,\mathsf{SNR}} \leq \sqrt[k+1]{\frac{(k+1)!\, g_{\mathsf{sd}} \prod_{i=1}^{k} g_{\mathsf{r_i d}} \prod_{i=1}^{k} g_{\mathsf{sr_i}}}{\prod_{i=1}^{k}(g_{\mathsf{sr_i}} + g_{\mathsf{r_i d}})}} \tag{64}$$

*Proof.* The max-flow min-cut bound shown in (63) implies that

$$\begin{aligned} P_{out_{k-relay}}(R) &= \mathbb{P}\{C_{k-relay}(h_{\mathsf{sr_1}},\ldots,h_{\mathsf{sr_k}},h_{\mathsf{sd}},h_{\mathsf{r_1 D}},\ldots,h_{\mathsf{r_k d}}) < R\} \\ &\geq \mathbb{P}\{(|h_{\mathsf{sd}}|^2 + \sum_{i=1}^{k} \min\{|h_{\mathsf{sr_i}}|^2, |h_{\mathsf{r_i d}}|^2\})\mathsf{SNR} < \mathsf{R}\} \\ &= \mathbb{P}\{|h_{\mathsf{sd}}|^2 + \sum_{i=1}^{k} \min\{|h_{\mathsf{sr_i}}|^2, |h_{\mathsf{r_i d}}|^2\} < \frac{R}{\mathsf{SNR}}\} \end{aligned}$$

Now by using lemmas 6.1 and 6.2 we have

$$\begin{aligned} \lim_{\substack{\mathsf{SNR}\to 0\\ \frac{R}{\mathsf{SNR}}\to 0}} \frac{p_{out_{k-relay}}(R)}{\left(\frac{R}{\mathsf{SNR}}\right)^{k+1}} &\geq \lim_{\substack{\mathsf{SNR}\to 0\\ \frac{R}{\mathsf{SNR}}\to 0}} \frac{\mathbb{P}\{|h_{\mathsf{sd}}|^2 + \sum_{i=1}^{k} \min\{|h_{\mathsf{sr_i}}|^2, |h_{\mathsf{r_i d}}|^2\} < \frac{R}{\mathsf{SNR}}\}}{\left(\frac{R}{\mathsf{SNR}}\right)^{k+1}} \\ &= \frac{\prod_{i=1}^{k}(g_{\mathsf{sr_i}} + g_{\mathsf{r_i d}})}{(k+1)!\, g_{\mathsf{sd}} \prod_{i=1}^{k} g_{\mathsf{r_i d}} \prod_{i=1}^{k} g_{\mathsf{sr_i}}} \end{aligned}$$

Thus

$$\lim_{\substack{\epsilon\to 0\\ \mathsf{SNR}\to 0}} \frac{C_{\epsilon_{k-relay}}}{\sqrt[k+1]{\epsilon}\,\mathsf{SNR}} \leq \sqrt[k+1]{\frac{(k+1)!\, g_{\mathsf{sd}} \prod_{i=1}^{k} g_{\mathsf{r_i d}} \prod_{i=1}^{k} g_{\mathsf{sr_i}}}{\prod_{i=1}^{k}(g_{\mathsf{sr_i}} + g_{\mathsf{r_i d}})}} \tag{65}$$

□

### 6.2.2 The Achievable Scheme: BAF protocol

As a consequence of the optimality of BAF protocol in the FD-relay channel we know that this protocol transfers the energy equal to

$$\min\{|h_{\mathsf{sr_i}}|^2, |h_{\mathsf{r_i d}}|^2\}\mathsf{SNR}$$

thorough each indirect path. Therefore in this case also BAF protocol will achieve the max-flow min-cut bound in the limit of low SNR and low outage probability and

$$\lim_{\substack{\epsilon\to 0\\ \mathsf{SNR}\to 0}} \frac{C_{\epsilon_{k-relay}}}{\sqrt[k+1]{\epsilon}\,\mathsf{SNR}} = \sqrt[k+1]{\frac{(k+1)!\, g_{\mathsf{sd}} \prod_{i=1}^{k} g_{\mathsf{r_i d}} \prod_{i=1}^{k} g_{\mathsf{sr_i}}}{\prod_{i=1}^{k}(g_{\mathsf{sr_i}} + g_{\mathsf{r_i d}})}} \tag{66}$$

# 7 Conclusion

In this paper we looked at the outage performance of the FD fading relay channel. We were able to find the $\epsilon$-outage capacity and the $\epsilon$-outage capacity per unit cost of this relay channel in the limit of low SNR and low probability of outage. We also showed that this optimal outage rate is achieved by Bursty Amplify-Forward protocol.

As the channel estimation is quite challenging in the low SNR regime we look at a non coherent scenario that neither the transmitter nor the receiver know the channel state. We showed that there is a scheme that uses bursty pulse position modulation for encoding and a type of energy detection for decoding and achieves the same rate as before (with the same outage probability). Hence the outage capacity of non coherent scenario is the same as the coherent scenario. We also investigate another extreme that the channel state information is available at both the transmitter and the receiver (full CSI). We show that this additional information will just slightly increase the outage capacity while the communication protocol gets quite complicated. The optimal scheme in this case is a combination of beam-forming and bursty amplify-forward protocols.

Finally we considered two important extensions in this paper. One is when we have an average sum power constraint on the source and the relay. We proposed the optimal power allocation to the source and the relay for this scenario. The other extension is when we can add more relays to help the source. Here we demonstrated how much the outage capacity increases by adding each additional relay.

# Acknowledgment

The authors would like to thanks Prof. Michael Gastpar for helpful comments and discussions.

# A Appendix: Some Probability Preliminaries

In this appendix we investigate some of the probabilities that we will need to analyze the communication protocols.

**Lemma A.1.** *Let $w = u + v$, where $u$ and $v$ are independent exponential random variables with mean $\mu_u$ and $\mu_v$, respectively. Then if $g(\epsilon)$ is a continuous function at $\epsilon = 0$ and $g(\epsilon) \to 0$ as $\epsilon \to 0$ we have*

$$\lim_{\epsilon \to 0} \frac{1}{g(\epsilon)^2} \mathbb{P}\{w < g(\epsilon)\} = \frac{1}{2\mu_u \mu_v} \tag{67}$$

*Proof.* The proof only requires basic probability calculations. □

**Lemma A.2.** *Let $\delta$ be positive, and let $r_\delta = \frac{vw}{v+w+\delta}$, where $v$ and $w$ are independent exponential random variables with mean $\mu_v$ and $\mu_w$, respectively. Let $h(\delta)$ be continuous with $h(\delta) \to 0$ as $\delta \to 0$. Then*

$$\lim_{\delta \to 0} \frac{1}{h(\delta)} \mathbb{P}\{r_\delta < h(\delta)\} = \mu_v^{-1} + \mu_w^{-1} \tag{68}$$

*Proof.* As $u$ and $v$ are exponential random variables with parameters $\lambda_v = \mu_v^{-1}$ and $\lambda_w = \mu_w^{-1}$, we have

$$\begin{aligned}
\mathbb{P}\{\frac{vw}{v+w+\delta} < h(\delta)\} &= \mathbb{P}\{vw < (v+w+\delta)h(\delta)\} \\
&= \mathbb{P}\{v(w-h(\delta)) < (w+\delta)h(\delta)\} \\
&= \mathbb{P}\{v < \frac{(w+\delta)h(\delta)}{w-h(\delta)} | w > h(\delta)\}\mathbb{P}\{w > h(\delta)\} + \mathbb{P}\{w < h(\delta)\} \\
&= \mathbb{P}\{v < \frac{(w'+\delta+h(\delta))h(\delta)}{w'}\} e^{-\lambda_w h(\delta)} + (1 - e^{-\lambda_w h(\delta)}) \\
&= \mathbb{P}\{v < h(\delta) + \frac{(\delta+h(\delta))h(\delta)}{w'}\} e^{-\lambda_w h(\delta)} + (1 - e^{-\lambda_w h(\delta)}) \\
&= e^{-\lambda_w h(\delta)} \int_0^\infty e^{-w'}(1 - e^{-\lambda_v (h(\delta) + \frac{(\delta+h(\delta))h(\delta)}{w'})}) dw' + (1 - e^{-\lambda_w h(\delta)}) \\
&= e^{-\lambda_w h(\delta)}(1 - e^{-\lambda_v h(\delta)} \int_0^\infty e^{-(w' + \frac{\lambda_v h(\delta)(\delta+h(\delta))}{w'})} dw') + (1 - e^{-\lambda_w h(\delta)}) \\
&= e^{-\lambda_w h(\delta)}(1 - e^{-\lambda_v h(\delta)} \int_0^\infty e^{-(w' + \frac{f(\delta)}{w'})} dw') + (1 - e^{-\lambda_w h(\delta)})
\end{aligned}$$

where $w' = w - h(\delta)$ and $f(\delta) = \lambda_v h(\delta)(\delta + h(\delta)) \to 0$ as $\delta \to 0$. But,

$$\int_0^\infty e^{\frac{-\beta}{4x} - \gamma x} dx = \sqrt{\frac{\beta}{\gamma}} K_1(\sqrt{\beta\gamma}) \tag{69}$$

where $K_1(.)$ is the modified Bessel function of the second type. So

$$\begin{aligned}
\lim_{\delta\to 0} \frac{1}{h(\delta)} \mathbb{P}\{r_\delta < h(\delta)\} &= \lim_{\delta\to 0} \left( \frac{e^{-\lambda_w h(\delta)}(1 - e^{-\lambda_v h(\delta)}[\sqrt{4f(\delta)} K_1(\sqrt{4f(\delta)})]) + 1 - e^{-\lambda_w h(\delta)}}{h(\delta)} \right) \\
&= \lim_{\delta\to 0} \frac{e^{-\lambda_w h(\delta)}(1 - e^{-\lambda_v h(\delta)})}{h(\delta)} + \lim_{\delta\to 0} \frac{1 - e^{-\lambda_w h(\delta)}}{h(\delta)} \\
&= \lambda_v + \lambda_w
\end{aligned}$$

the second equality is true as $\lim_{\epsilon\to 0} \epsilon K_1(\epsilon) = 1$.

□

**Lemma A.3.** *Let $u$, $v$ and $w$ be independent exponential random variables with mean $\mu_u$, $\mu_v$, and $\mu_w$ respectively. Let $\epsilon$ be positive and let $g(\epsilon) > 0$ be continuous with $g(\epsilon) \to 0$ as $\epsilon \to 0$. Then*

$$\lim_{\epsilon\to 0} \frac{1}{g^2(\epsilon)} \mathbb{P}\{u + \frac{vw}{v + w + \epsilon} < g(\epsilon)\} = \frac{\mu_u^{-1}(\mu_v^{-1} + \mu_w^{-1})}{2}. \tag{70}$$

*Proof.* As $u$, $v$ and $v$ are exponential random variables with parameters $\lambda_u = \mu_u^{-1}$, $\lambda_v = \mu_v^{-1}$ and $\lambda_w = \mu_w^{-1}$ respectively, then using the same $r_\epsilon$ as in Lemma A.2 we have

$$\begin{aligned}
\mathbb{P}\{u + \frac{vw}{v + w + \epsilon} < g(\epsilon)\} &= \mathbb{P}\{u + r_\epsilon < g(\epsilon)\} \\
&= \int_0^{g(\epsilon)} \mathbb{P}\{r_\epsilon < g(\epsilon) - u\} p_u(u) du \\
&= g(\epsilon) \int_0^1 \mathbb{P}\{r_\epsilon < g(\epsilon)(1 - u')\} \lambda_u e^{-\lambda_u g(\epsilon) u'} du' \\
&= g(\epsilon)^2 \int_0^1 (1 - u') [\frac{\mathbb{P}\{r_\epsilon < g(\epsilon)(1 - u')\}}{g(\epsilon)(1 - u')}] \lambda_u e^{-\lambda_u g(\epsilon) u'} du'
\end{aligned}$$

where in the third equality we have used the change of variables $u' = u/g(\epsilon)$. But by Lemma A.2 with $\delta = \epsilon$ and $h(\delta) = g(\epsilon)(1 - u')$, the quantity in the brackets approaches $\lambda_v + \lambda_w$ as $\epsilon \to 0$ and gives the result. □